\newcommand{\Xcomment}[1]{}

\Xcomment{
\documentclass[twoside,leqno,twocolumn]{article}
\usepackage{ltexpprt}
}

\documentclass[12pt]{article}
\usepackage{setspace}
\newtheorem{lemma}{Lemma}
\newtheorem{theorem}{Theorem}
\newtheorem{corollary}{Corollary}
\newcommand{\proof}{\noindent \emph{Proof.}\ }
\newcommand{\proofend}{$\Box$\\}

\Xcomment{
JOURNAL VERSION:
\begin{itemize}
\item Insert appendices;
\item Insert alternative analysis in Section 3.2 (time not cost);
\item Can the proof of Lemma 2 be restated as a potential argument?
\item Leaf deletes?
\item Definition of $n$?
\end{itemize}
}

\usepackage{epsfig}
\usepackage{latexsym}
\newcommand{\rank}{\mathit{rank}}

\newcommand{\size}{\mathit{size}}
\newcommand{\merge}{\mathit{merge}}

\newcommand{\nca}{\mathit{nca}}
\newcommand{\topn}{\mathit{top}}
\newcommand{\nul}{\mathit{null}}
\newcommand{\labeln}{\mathit{\ell}}
\newcommand{\rootn}{\mathit{root}}
\newcommand{\bottom}{\mathit{bottom}}

\newcommand{\treemin}{\mathit{treemin}}
\newcommand{\pathmin}{\mathit{pathmin}}

\begin{document}

\title{\Large Data Structures for Mergeable Trees\footnote{A preliminary version of some of this material appeared in the conference paper ``Design of data structures for mergeable trees'', \emph{Proceedings of the 17th Annual ACM-SIAM Symposium on Discrete Algorithms}, pages 394--403, 2006.}}
\author{Loukas Georgiadis$^{1}$ \and Haim Kaplan$^{2}$ \and Nira Shafrir$^{2}$ \and Robert E. Tarjan$^{3}$ \and Renato F. Werneck$^{4}$}
\date{\today}

\maketitle


\begin{abstract} \small\baselineskip=9pt

Motivated by an application in computational topology, we consider a novel variant of the problem of efficiently maintaining dynamic rooted trees.  This variant requires merging two paths in a single operation.  In contrast to the standard problem, in which only one tree arc changes at a time, a single merge operation can change many arcs.  In spite of this, we develop a data structure that supports merges on an $n$-node forest in $O(\log^2 n)$ amortized time and all other standard tree operations in $O(\log n)$ time (amortized,  worst-case, or randomized depending on the underlying data structure).  For the special case that occurs in the motivating application, in which arbitrary arc deletions (\emph{cuts}) are not allowed, we give a data structure with an $O(\log n)$ time bound per operation.  This is asymptotically optimal under certain assumptions.  For the even-more special case in which both cuts and parent queries are disallowed, we give an alternative $O(\log n)$-time solution that uses standard dynamic trees as a black box. This solution also applies to the motivating application.  Our methods use previous work on dynamic trees in various ways, but the analysis of each algorithm requires novel ideas.  We also investigate lower bounds for the problem under various assumptions.
\end{abstract}

\footnotetext[1]{Hewlett-Packard Laboratories, Palo Alto, CA, 94304.
Part of this work was done while this author was at Princeton University. E-mail: {\tt loukas.georgiadis@hp.com}.}
\footnotetext[2]{Tel-Aviv University, Tel-Aviv, Israel. E-mail: {\tt haimk@math.tau.ac.il} and {\tt shafrirn@post.tau.ac.il}.}
\footnotetext[3]{Department of Computer Science, Princeton University, 35 Olden Street, Princeton, NJ 08540 and Hewlett-Packard
Laboratories, Palo Alto, CA, 94304. E-mail: {\tt ret@cs.princeton.edu}.}
\footnotetext[4]{Microsoft Research Silicon Valley, 1065 La Avenida, Mountain View, CA, 94043. Part of this work
was done while this author was at Princeton University. E-mail: {\tt renatow@microsoft.com}.

\vspace{.1cm}
\noindent Research by Loukas Georgiadis, Robert E. Tarjan, and Renato F. Werneck at Princeton University was partially supported by the Aladdin Project, NSF Grant No 112-0188-1234-12. Work by Haim Kaplan and Nira Shafrir was partially supported by Grant 975/06 from the Israel Science Foundation (ISF).}

\thispagestyle{empty}

\setcounter{page}{1}
\section{Introduction}
\label{sec:intro}

\begin{figure}[h]
\addtolength{\abovecaptionskip}{-.5cm}
\begin{center}
\resizebox{1\textwidth}{!} {\includegraphics{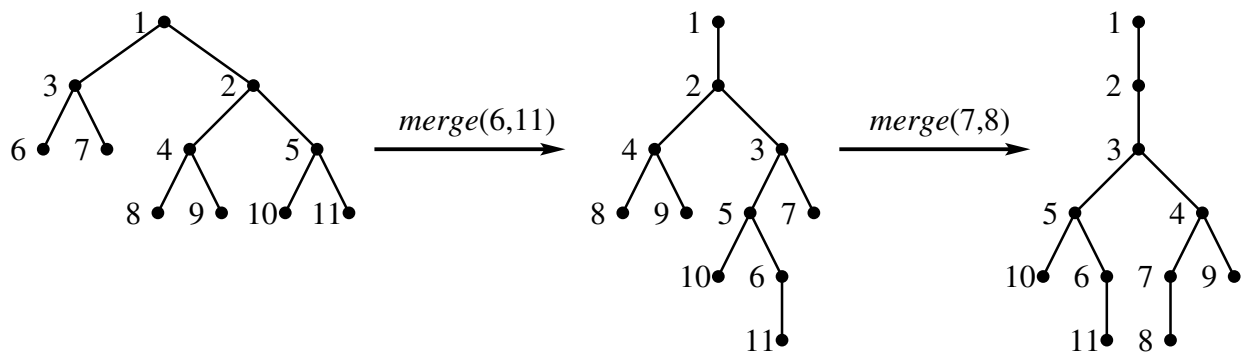}}
\end{center}
\caption{\label{fig:examples} Two successive merges. The nodes are
identified by label.}
\end{figure}

A \emph{heap-ordered forest} is a set of node-disjoint rooted trees, in which each node $v$ has a real-valued \emph{label} $\labeln(v)$, and the labels are in heap order: if $p(v)$ is the parent of $v$, $\labeln(v) \ge \labeln(p(v))$.  We consider the problem of maintaining a heap-ordered forest, initially empty, subject to an arbitrary intermixed sequence of the following kinds of operations:

\begin{itemize}
\vspace{-.75mm}
\addtolength{\itemsep}{-1.75mm}
\item {\em parent}$(v)$: Return the parent $p(v)$ of $v$, or null if $v$ is a tree root.
\item {\em root}$(v)$: Return the root of the tree containing $v$.
\item {\em nca}$(v,w)$: Return the nearest common ancestor of $v$ and $w$, or null if $v$ and $w$ are in different trees.
\item {\em insert}$(v, x)$: Create a new, one-node tree consisting of node $v$ with label $x$; $v$ must be in no other tree.
\item {\em link}$(v,w)$: Make $w$ the parent of $v$ by adding the arc $(v, w)$; $v$ must be a root, $w$ must be in another tree,
and $\labeln(v) \ge \labeln(w)$.
\item {\em cut}$(v)$: Delete the arc from $v$ to its parent, making $v$ a root; do nothing if $v$ is already a root.
\item {\em delete}$(v)$: Delete $v$ from the forest; $v$ must be a leaf (a node with no children).
\item {\em merge}$(v,w)$: Let $P$ and $Q$, respectively, be the paths from $v$ and $w$ to the roots of their respective trees.  Restructure the tree or trees containing $v$ and $w$ by merging the paths $P$ and $Q$ while preserving heap order. See Figure~\ref{fig:examples}.
\end{itemize}

This is the \emph{mergeable trees problem}.  This problem arises in an algorithm of Agarwal et al.~\cite{AEHW04,AEHW06} that computes the structure of 2-manifolds embedded in $\mathcal{R}^3$.  In this application, the tree nodes are the critical points of the manifold (local minima, local maxima, and saddle points), with labels equal to their heights.  The algorithm computes the critical points and their heights during a sweep of the manifold, and pairs up the critical points into so-called \emph{critical pairs} using mergeable tree operations.  This use of mergeable trees is actually a special case: there are no cuts.  As we shall see, one can also avoid parent operations, by changing the pairing process to do two sweeps, one upward and one downward, instead of a single sweep.

The mergeable trees problem is a new variant of the well-studied \emph{dynamic trees} problem, which calls for the maintenance of a forest of trees subject to all the mergeable tree operations except merge.  Nodes are not heap-ordered by label; instead, each node or arc has an arbitrary associated value, and values can be accessed or changed one node or arc at a time, an entire path at a time, or even an entire tree at a time.  The original use of dynamic trees was in a network flow algorithm~\cite{ST83}.  In that application, each arc has an associated real value, its residual capacity.  The maximum value on a path can be computed in a single operation, and a given value can be subtracted from all arcs on a path in a single operation.

There are several versions of the dynamic trees problem that differ in what kinds of values are allowed, whether values can be combined over paths or over entire trees (or both) at a time, whether the trees are unrooted, rooted, or ordered (each set of siblings is ordered), and exactly what operations are allowed.  For all these versions of the problem, there are algorithms that perform a sequence of tree operations in logarithmic time per operation,  either amortized~\cite{ST85,TW05}, worst-case~\cite{AHTdL05,Fre85,ST83}, or randomized~\cite{ABHVW04}. The nca operation is not completely standard for dynamic trees; it accesses two paths rather than one.  But it is easy to extend any of the efficient implementations of dynamic trees to support nca in $O(\log n)$ time.  Indeed, Sleator and Tarjan~\cite{ST83} and Alstrup et al.~\cite{AHTdL05} explicitly describe how to do this.

The main novelty, and the main difficulty, in the mergeable trees problem is the merge operation.  Although dynamic trees support global operations on node and arc values, the underlying trees change only one arc at a time, by links and cuts (and deletions, which are in effect cuts).  In contrast, a merge operation can delete and add many arcs, even a linear number, simultaneously.  Nevertheless, there are efficient implementations of mergeable trees.  We give three.  In Section \ref{sec:dyntrees} we show that the amortized number of arcs changed by a merge is logarithmic.  This allows us to implement mergeable trees using dynamic trees directly, with a logarithmic (or better) time bound for every operation except merge, and a log-squared amortized bound for merge.  In Section \ref{sec:part-rank} we consider the special case in which there are no cuts.  For this case we combine ideas in a previous implementation of dynamic trees with a novel analysis, to obtain an algorithm with a logarithmic or better time bound for every operation.  In Section \ref{sec:weak} we consider the special case in which there are neither cuts nor parent operations.  For this case we give an alternative logarithmic-time solution that represents mergeable trees implicitly as dynamic trees: a merge becomes either a link, or a cut followed by a link.  Either of the methods of Sections \ref{sec:part-rank} and \ref{sec:weak} can be used to efficiently pair critical points.  We discuss this application in Section \ref{sec:pairing}, including filling in a gap in the pairing algorithm of Agarwal et al.~\cite{AEHW04,AEHW06}.  In Section \ref{sec:complexity} we discuss lower bounds and related issues for various forms of the mergeable trees problem.

In discussing the mergeable trees problem we shall use the following terminology.  Each tree arc is directed from child to parent, so that a path leads from a node toward the root of its tree, a direction we call \emph{upward}.  Node $v$ is a \emph{descendant} of $w$, and $w$ is an \emph{ancestor} of $v$, if the path from $v$ to $\rootn(v)$ contains $w$.  (This includes the case $v$ = $w$.)  We also say $v$ is \emph{below} $w$, and $w$ is \emph{above} $v$.  If $v$ is neither an ancestor nor a descendant of $w$, then $v$ and $w$ are \emph{unrelated}. We denote by $\size(v)$ the number of descendants of $v$, including $v$.  We denote by $P[v, w]$ the path from node $v$ to node $w$, and by $P[v, w)$ and $P(v, w]$, respectively, the subpath of $P[v, w]$ obtained by deleting $w$ or deleting $v$; if $v = w$, $P[v, w)$ and $P(v, w]$ are empty.  By extension, $P[v, \nul) = P[v, \rootn(v)]$.  We denote by $\bottom(P)$ and $\topn(P)$ the first (bottommost) and last (topmost) nodes on a path $P$, and by $|P|$ the number of nodes on $P$.

We replace each operation $\mathit{link}(v, w)$ by $\merge(v, w)$, since they have the same effect.  This avoids the need to consider link explicitly as a mergeable tree operation.  We denote by $m$ the number of merges, including those replacing links. We denote by $n$ the number of inserts of nodes that are eventually in trees that participate in merges; any node that does not participate remains in a single-node tree, on which all operations take constant time. The definition of $n$ implies that $n \le 2m$. All our data structures take space linear in the number of insertions.

In a merge, the merge order is unique if all nodes have distinct labels.  If not, we can break ties using node identifiers.  To simplify things, and without loss of generality, we shall assume that the node labels are the nodes themselves, and that the nodes that are in trees that eventually participate in merges are the integers $1$ through $n$, numbered in label order.  We treat a node that is deleted and reinserted as an entirely new node when it is reinserted, with a new number. We also treat null as being less than any node.

This paper is a major reworking of a conference paper~\cite{GTW06}.  We have simplified the analysis of the algorithms in Sections \ref{sec:dyntrees} and \ref{sec:part-rank}, added a detailed description of the critical point pairing application (Section \ref{sec:pairing}), and added the algorithm in Section \ref{sec:weak}, which is the contribution of the two new authors (Kaplan and Shafrir).

\section{Mergeable Trees as Dynamic Trees}
\label{sec:dyntrees}

In this section we explore the obvious way to implement mergeable trees, which is to represent them by dynamic trees of exactly the same structure.  Then the mergeable tree operations parent, root, nca, insert, cut, and delete become exactly the same operations on dynamic trees.  In order to do merges, we need one additional operation on heap-ordered dynamic trees:

\begin{itemize}
\item $\mathit{topmost}(v,w)$: Return the smallest (topmost) ancestor of $v$ that is strictly greater than $w$, assuming $v>w$.
\end{itemize}

This operation accesses the path $P[v, \nul)$.  It is easy to extend any of the efficient implementations of dynamic trees to support topmost in $O(\log n)$ time.

\begin{figure}
\addtolength{\abovecaptionskip}{-.5cm}
\begin{center}
\resizebox{0.85\textwidth}{!} {\includegraphics{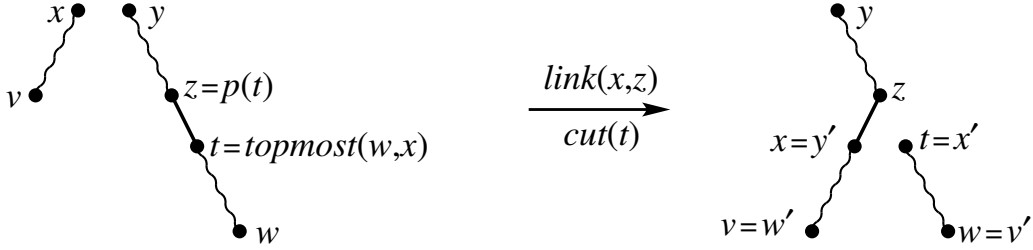}}\\
\end{center}
\caption{\label{fig:topmost} A merge step. Straight lines are arcs, wavy lines are tree paths. Primed variables are values after the step.}
\end{figure}

To perform $\merge(v, w)$, begin by computing $u = \nca(v, w)$.  Stop if $u = v$ or $u = w$.  Otherwise, walk down the paths $P[v, u)$ and $P[w, u)$ toward $v$ and $w$, merging them step-by-step.  Maintain two current nodes $x$ and $y$, initially $\mathit{topmost}(v, u)$ and $\mathit{topmost}(w, u)$, respectively.  If $x < y$, swap $x$ and $y$ and $v$ and $w$, respectively.  If $u \neq \nul$, do $\mathit{cut}(x)$.  While $x < w$, repeat the following step:

\begin{description}
\item[Merge Step:] Let $t = \mathit{topmost}(w, x)$.  Do $\mathit{link}(x, \mathit{parent}(t))$ and $\mathit{cut}(t)$. Set $y$ equal to $x$, set $x$ equal to $t$, and swap $v$ and $w$. (See Figure \ref{fig:topmost}.)
\end{description}

To finish the merge, do $\mathit{link}(x, w)$.

After the merge initialization, $x$ and $y$ are the children of $u$ that are ancestors of $v$ and $w$, respectively, or $\mathit{root}(v)$ and $\mathit{root}(w)$ if $u$ is null.  At the beginning of a merge step,  $x$ and $y$ are the tops of the paths remaining to be merged, $x$ is a root, and $y < x < w$.  The merge step finds the node $t$ whose parent is the parent of $x$ after the merge and updates values appropriately.  The correctness of the merging algorithm follows.

Each merge step takes a constant number of dynamic tree operations, as do the initialization and finalization.  There is one parent change per merge step plus one per merge.  We shall obtain an $O(\log n)$ bound on the amortized number of parent changes per merge, which implies an $O(\log^2 n)$ amortized time bound for merging, assuming that the underlying dynamic tree data structure has an $O(\log n)$ time bound per operation.

\begin{lemma}
\label{lemma:dyn-trees}
The total number of parent changes over all merges is $O(m \log n)$.
\end{lemma}
\proof We use an amortized analysis~\cite{Tar85}.  Each state of the data structure has a non-negative potential; the initial, empty structure has a potential of zero.  We define the cost of an operation to be the number of parent changes it causes; we define the amortized cost of an operation to be its cost plus the net decrease in potential it causes.  Then the sum of the amortized costs of all the operations is an upper bound on the total number of parent changes caused by all the operations.

With each arc $(v, w)$ we associate $2 \lg (v - w) + 1$ units of potential, where $\lg$ is the base-two logarithm. Of this amount, we assign $1$ to $(v,w)$, $\lg (v-w)$ to $v$, and $\lg (v-w)$ to $w$.  Thus each node has potential associated with its parent (if it has a parent) and with each of its children.  We call the former its \emph{parent potential} and the sum of the latter its \emph{child potential}.  The total potential is the sum of the potentials.

The only operations that affect the structure of the forest are merges, cuts, and node deletions.  A cut or node deletion creates at most one new (null) parent and decreases the potential by at least one, so its amortized cost is non-positive. Consider a merge. If $u = \nul$, the initial link of $x$ increases the potential by at most $2\lg n + 1$. Every other change in a node potential is non-positive. Consider a merge step. Let $p(t)$ be the parent of $t$ before the cut of $t$ occurs, and let $p'(t)$ be the new parent of $t$, which it acquires either in the next merge step, or at the end of the merge if this is the last step.  Then $t > p'(t) \ge x > p(t)$.  If $t - p'(t) \le (t - p(t))/2$, then the parent potential of $t$ decreases by at least one as a result of its parent changing.  If $x - p(t) \le (t - p(t))/2$, then the child potential of $p(t)$ decreases by at least one as a result of $p(t)$ losing $t$ as a child but gaining $x$.  One of these two cases must occur. The amortized cost of a merge is thus $O(\log n)$ (for the initial link of $x$ if $u = \nul$) plus a net of at most zero per merge step (one parent creation minus at least one unit of potential) plus one (for one extra parent creation per merge).
\proofend

If we use any of the several implementations of dynamic trees that support all operations in $O(\log n)$ time, Lemma \ref{lemma:dyn-trees} gives an $O(\log^2 n)$ amortized time bound for merge; all the other operations have the same time bound as in the underlying dynamic tree structure.  With this method one can maintain parent pointers explicitly, which makes the worst-case time for the parent operation $O(1)$.

The proof of Lemma \ref{lemma:dyn-trees} gives something a little stronger: the amortized cost of a merge is $O(1)$ unless the merge combines two trees.  If there are no cuts, then the number of merges that can combine two trees is at most $n - 1$,
which means that the total number of parent changes is $O(m + n \log n)$. This bound is tight, as we show in Section \ref{sec:complexity}.  The total time for merges becomes $O(m \log n + n \log^2 n)$.  In the absence of cuts, we can get an even better bound on the merge time by changing the algorithm, as we shall see in the next two sections.

In a preliminary version of their paper~\cite{AEHW04}, Agarwal et al. proposed representing mergeable trees by dynamic trees as we do here, but they suggested a different merging algorithm, in which the nodes of the shorter merge path are inserted one-by-one into the longer merge path.  Although they claimed an $O(n \log n)$ bound on the total number of node insertions (assuming no cuts), this bound is incorrect: the worst-case number of node insertions in the absence of cuts is $\Omega(n^{3/2})$, as we show in Section \ref{sec:complexity}.  Thus this method of merging does not give even a polylogarithmic amortized bound for merge.

\section{Mergeable Trees via Partition by Rank}
\label{sec:part-rank}

In this and the next section we develop two different methods to achieve an $O(\log n)$ bound per merge, if there are no cuts.  For the moment we also ignore leaf deletions; we discuss how to handle them at the end of the section.  Our first method uses an idea from Sleator and Tarjan's~\cite{ST83, ST85} implementation of dynamic trees: we partition each tree into node-disjoint paths, and implement the various tree operations as appropriate sequences of path operations.  The updates we need on paths are deletions of top nodes and arbitrary insertions of single nodes.  We also need a variant of the topmost query defined in Section \ref{sec:dyntrees}.

\begin{figure} [htb]
\addtolength{\abovecaptionskip}{-.5cm}
\begin{center}
\resizebox{0.5\textwidth}{!} {\includegraphics{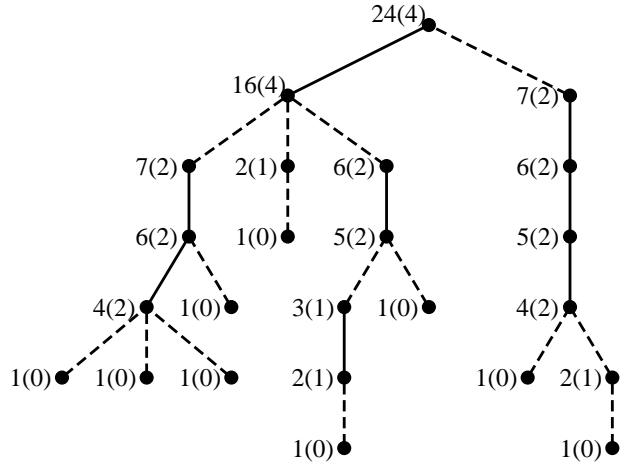}}
\end{center}
\caption{\label{fig:solid} A tree partitioned by rank into solid
paths, with the corresponding sizes and, in parentheses, ranks.}
\end{figure}

We define the \emph{rank} of a node $v$ to be $\lfloor \lg \size(v)\rfloor$.  Ranks are integers in the range from zero to $\lg n$.  We
decompose the forest into solid paths by defining an arc $(v, w)$ to be \emph{solid} if $\rank(v) = \rank(w)$ and \emph{dashed} otherwise. Since a node can have at most one solid arc from a child, the solid arcs partition the forest into node-disjoint solid paths. See Figure \ref{fig:solid}.  Our path partition is a variant of one used by Sleator and Tarjan~\cite{ST83}: theirs makes an arc $(v, w)$ solid if $\size(v) > \size(w)/2$;  our solid arcs are a subset of theirs. We call a node a \emph{top node} if it is the top of its solid path. We call a non-root node a \emph{solid child} if its arc to its parent is solid and a \emph{dashed child} otherwise.

The merging algorithm uses the same approach as in Section \ref{sec:dyntrees}: to merge $v$ and $w$, first ascend the paths from $v$ and $w$ to find their nearest common ancestor $u$, then merge the traversed paths top-down. Each of the merge paths is a sequence of parts of solid paths.  An added complication in merging is that we must update the path partition, which requires keeping track of ranks.  In the absence of cuts, no node can ever decrease in rank, and we can charge the work of merging against rank changes.  By using an appropriate form of binary search tree to represent solid paths, we can obtain a logarithmic amortized time bound for merging.

The remainder of this section develops and analyzes this method. Section \ref{sec:nca} discusses the access operations, parent, root, and nca. Section \ref{sec:merge} describes the merging algorithm. Section \ref{sec:merge-analysis} analyzes the running time of merging. Section \ref{sec:merge-representation} discusses the use of search trees to represent solid paths and completes the analysis of merging. Section \ref{sec:deletions} describes how to extend the method to support leaf deletions.

\subsection{Access Operations}
\label{sec:nca}

We represent the path partition using three sets of pointers and a set of headers, one for each path.  Each node has pointers to its parent (null if it is a root) and to its solid child (null if it has none).  This makes each solid path a doubly-linked list, and allows accessing the parent or solid child of a node in constant time.  Efficient computation of roots and of nearest common ancestors requires fast access from any node $v$ to $\topn(v)$, the top of the solid path containing $v$.  To allow such access while also allowing fast updating, we use one level of indirection, which is the purpose of the path headers:  each node points to the header of its path; the header points to the top of the path.

Both $\rootn(v)$ and $\nca(v, w)$ take $O(\log n)$ worst-case time.  To do $\rootn(v)$, traverse the path from $v$ to the root, step-by-step.  A step is from a node to its parent if it is a top node or to the top of its solid path if it is not. Each such step takes constant time via either a parent pointer or a path header.  The traversal reaches a new solid path, of higher-rank nodes, in at most two steps, and thus reaches the root in at most $2\lg n$ steps.  The nca operation is similar but requires traversing two paths concurrently.  To do $\nca(v, w)$, traverse the paths from $v$ and $w$ bottom-up, taking the next step from the larger of the two current nodes, and stop when reaching a common solid path or reaching two roots.  If $x$ and $y$ are the last nodes reached by the concurrent traversals, the nearest common ancestor of $v$ and $w$ is $\min \{x, y\}$ if $x$ and $y$ are on a common solid path, null otherwise.  The concurrent traversal reaches a common solid path or a pair of roots in at most $4\lg n$ steps.

\subsection{Merging}
\label{sec:merge}

Merging requires the ability to keep track of ranks, which we do by keeping track of sizes.  To make this efficient, we store explicitly
only the sizes of top nodes.  Since we can access a top node from any node on its solid path in constant time, and since all nodes on a solid path have the same rank, we can compute the rank of any node in constant time.  To help maintain the sizes of top nodes, we also store with every node $x$ its \emph{dashed size} $d(x)$, defined to be one (to count $x$ itself) plus the sum of the sizes of the dashed children of $x$. We can compute the size of any solid child $x$ from that of its parent in constant time using the following equation:
\begin{equation}
\label{eq:size}
\size(x) = \size(p(x)) - d(p(x)).
\end{equation}

Merging uses the following variant of the topmost query:

\begin{itemize}
\item $\mathit{topmost}(v, w)$: Return the topmost node on the solid path containing $v$ that is strictly greater than $w$, or null if there is no such node.
\end{itemize}

To perform $\merge(v, w)$, begin by computing the nearest common ancestor $u$ of $v$ and $w$ by the method of Section \ref{sec:nca}, keeping track of the two sequences of nodes visited by the traversals from $v$ and $w$. Stop if $u=v$ or $u=w$. Otherwise, traverse the paths to $u$ from $v$ and $w$ top-down, merging them step-by-step, updating solid paths as necessary. To do this, maintain two current nodes $x$ and $y$, initially the children of $u$ that are ancestors of $v$ and $w$, respectively, or $\rootn(v)$ and $\rootn(w)$ if $u = \nul$.  If $x < y$, swap $x$ and $y$ and $v$ and $w$, respectively.  While $x < w$, repeat the following step:
\begin{description}
\item[Merge Step:] Let $s$ be the first (bottommost) node on the solid path containing $y$ that was reached during the traversal from $w$.  If $x < s$, let $t = \mathit{topmost}(y, x)$; otherwise, let $t$ be the node below $s$ that was reached during the traversal from $w$.  (If $x > s$, such a $t$ exists because $x < w$.)  Make $p(t)$ the parent of $x$. (Node $p(t)$ is a descendant of $y$.)  Update the solid paths that change as a result of this parent change.  Set $y$ equal to $t$.  If $x < y$, swap $x$ and $y$ and $v$ and $w$, respectively.
\end{description}
To finish the merge, make $w$ the parent of $x$ and update the solid paths accordingly.\\

This algorithm is like the merging algorithm in Section \ref{sec:dyntrees}, but it can make more parent changes, because it proceeds one solid path at a time.  Such extra changes occur only in merge steps for which $x > s$, which proceed from one solid path to another without doing a topmost query.  If $x < s$, $t = \mathit{topmost}(y, x)$ is on the solid path containing $s$ and $y$, since $x > y$.

Some details of the algorithm remain to be filled in.  To compute the initial values of $x$ and $y$ if $u \neq \nul$, let $s$ and $t$, respectively, be the last nodes on the paths traversed from $v$ and $w$, respectively, other than $u$. If $\rank(s) < \rank(u)$, then $x = \topn(s)$; otherwise, $x$ is the solid child of $u$.  Similarly, if $\rank(t) < \rank(u)$, then $y = \topn(t)$; otherwise, $y$ is the solid child of $u$.

We also need to update solid paths.  In our description of how to do this, primes denote updated values.  Let $z = p'(x)$: in a merge step, $z = p(t)$; in the finalization of a merge, $z = w$.  The only nodes whose rank can change are ancestors of $z$ on the same solid path as $z$.  Apply the appropriate one of the following two cases:

\begin{description}
\item[Case 1:] Node $x$ is a solid child.  Then $y$ is a dashed child, and all nodes on $P[z, y]$ change rank to $\rank(x) = \rank(p(x))$.  Make $z$ the parent of $x$, make $y$ the solid child of $p(y)$, and set $d'(p(y)) = d(p(y)) - \size(y)$.  If $z$ has a solid child $c$, compute its size by walking down along the path $P[c, y]$ applying equation (\ref{eq:size}), and then set $d'(y) = d(y) + \size(c)$.  This makes $c$ a dashed child.   Make $x$ the new solid child of $z$.  Change the header of every node on $P[z, y]$ to that of $x$, and make the header of $c$ (if c exists) point to $c$. 

\item[Case 2:] Node $x$ is not a solid child.  Compute the old rank of $z$.   Make $x$ a dashed child of $z$ by setting $d'(z) = d(z) + \size(x)$ and either setting $size'(\topn(z)) = \size(\topn(z)) + \size(x)$ if $x$ is a root (this only occurs at the beginning of a merge of two nodes in different trees) or setting $d'(p(y)) = d(p(y)) - \size(x)$ if $x$ is not a root.  Make $z$ the parent of $x$.  Now all values are correct for the current path partition, but the partition is not by rank.  To correct this, find the nodes that change rank by starting at $\topn(z)$ and walking down the solid path computing the new size (using equation (\ref{eq:size})) and rank of each node reached, until reaching a node whose rank does not change or walking off the bottom of the path.

    This identifies the nodes whose rank changes and the arcs that change type.  There are at most three such arcs: $(y, p(y))$ can become solid; $(x, z)$ can become solid; and either the arc to $z$ from its old solid child if it has one or an arc on the path $P[z, \topn(z)]$ can become dashed.  This follows from an examination of three cases.  If $\rank(x) < \rank(z)$, then zero or more nodes on $P[z, \topn(z)]$ increase in rank by one.  If $\rank(x) = \rank(z)$, then all nodes on $P[z, \topn(z)]$ increase in rank to $\rank(x) + 1$.  If $\rank(x) > \rank(z)$, then all nodes on $P[z, \topn(z)]$ increase in rank to $\rank(x)$ or $\rank(x) + 1$.  In each case at most one arc becomes dashed.

    In addition to updating solid child pointers, change a dashed arc $(a, b)$ to solid or vice-versa by subtracting or adding the new size of $a$ to the dashed size of $b$, respectively: if $(a, b)$ becomes dashed, $b$ increases in rank, and $\size'(a)$, which is needed both to update the dashed size of $b$ and since $a$ becomes a top node, is computed during the walk down the solid path.  Update the headers as follows.  If $\topn(z)$ is not a root, make the header of $p(\topn(z))$ also the header of each node whose rank increases to $\rank(p(\topn(z))$.  (In this case $\rank(\topn(p(z)) > \rank(z)$.)  For each node whose rank increases to $\rank(x)$, change its header to that of $x$.  If $(a, b)$ becomes dashed, make the header of $a$ point to $a$.  If one or more nodes have their headers change to that of $x$, make the header of $x$ point to the topmost such node. 
\end{description}

This completes the description of merging and of the data structure, except for the implementation of topmost queries.  To make these queries efficient, we represent each solid path by a suitable kind of search tree (in addition to parent and solid child pointers and headers).  In addition to topmost queries, this structure supports insertion of a node into a solid path above or below a given node, and deletion of the top of a solid path.  We do such insertions and deletions during Cases 1 and 2, as follows.  Walk down along the solid path from $\topn(z)$.  Delete from this solid path each node whose rank changes, and insert the node either above $x$ (if its new rank is that of $x$) or below its parent (in Case 2 if $\topn(z)$ is not a root and the new rank of the node is the rank of its parent).

\subsection{Analysis of Merging}
\label{sec:merge-analysis}

In this section we analyze the running time of merging, independent of the type of search tree used to represent solid paths.  We use this analysis in the next section to choose search trees that give an $O(\log n)$ amortized time for merging.

\begin{lemma}
\label{lemma:number-of-steps} The total number of solid path insertions and deletions is $O(n \log n)$.  The total number of merge steps is $O(m \log n)$.
\end{lemma}
\proof
Since there are no cuts, no node can decrease in rank, and the total number of increases in rank is at most $n \lg n$.  Each solid path deletion or insertion is of a node whose rank increases, so there are at most $n \lg n$ of each.  There are $O(n \log n)$ merge steps that cause a rank increase.  A merge step that does not cause a rank increase must result in Case 2 with $\rank(x) < \rank(p(t)) = \rank(y)$.  Consider how $r = \max \{ \rank(x), \rank(y) \}$ changes as the result of a merge step.  The value of $r$ is between 0 and $\lg n$ and cannot increase.  A merge step that does not cause a rank increase is either the last step of the merge, or decreases $r$ by at least one (if $x > s$), or is followed by a merge step resulting in Case 1 and hence causing a rank increase (if $x < s$).  It follows that the number of merge steps that do not cause a rank increase, and hence the total number of merge steps, is $O(m \log n)$. \proofend

\begin{corollary}
\label{cor:number-of-steps} The total time for all merges is $O(m \log n)$ plus the time for $O(m \log n)$ topmost queries and $O(n \log n)$ solid path insertions and deletions.
\end{corollary}
\proof There is at most one topmost query per merge step.  Not counting the time for topmost queries and solid path insertions and deletions, the time for a merge step is $O(1)$ plus $O(1)$ per solid path deletion.  The bound follows from Lemma \ref{lemma:number-of-steps}. \proofend

By Corollary \ref{cor:number-of-steps}, the amortized time for a merge is $O(\log n)$, not counting the time for topmost queries and solid path insertions and deletions.  If we represent each solid path by a binary search tree such as a red-black tree~\cite{GS78} or a splay tree~\cite{ST85}, then the time for an insertion, deletion, or topmost query is $O(\log n)$, giving the same $O(\log^2 n)$ amortized bound for merging as in Section \ref{sec:dyntrees}.  To obtain a better bound, we need a more-refined analysis of the topmost queries and we use a more-sophisticated kind of search tree to represent solid paths.

We define the \emph{cost} of a topmost query $t = \mathit{topmost}(y,x)$ to be $\lg |P[p(t),y]|$.

\begin{lemma}
\label{lemma:topmost}
The total cost of all the topmost queries over all merges is $O(n \log n)$.
\end{lemma}
\proof
We estimate the cost of two types of topmost queries separately.  We call a query \emph{type-1} if it occurs in a merge step such that $\rank(x) \ge \rank(y)$ and \emph{type-2} otherwise.  The total cost of the type-1 queries is easy to bound.  The parent change after a type-1 query causes every node on $P[p(t), y]$ to increase in rank.  It follows that the total cost of type-1 queries is at most $n \lg n$.

The parent change after a type-2 query does not necessarily cause the nodes on $P[p(t), y]$ to increase in rank, but it does cause their sizes to increase, and by analyzing these size increases we can get a bound on the total cost of type-2 queries.  Consider a type-2 query.  Let $d = |P[p(t), y]|$ and $k = \rank(y) - \rank(x)$.  The cost of the query is $\lg d$, of which we charge $1/2^k$ to each node on $P[p(t), y]$ and the residue to $x$.  The residual charge to $x$ is
$$
\lg d - d/2^k \le \lg d - 2^{(\lg d - k)} < \lg d - \lg d + k = k.
$$
We complete the proof by showing that the total charge over all type-2 queries is $O(n \log n)$.  When $1/2^k$ is charged to a node $b$ on $P[p(t), y]$, the size of $b$ increases by at least $2^{(rank(y) - k)} = 2^{(rank(b) - k)}$.  While at a given rank, such a node $b$ accumulates less than one unit of total charge, because if it accumulates one or more units of charge, its size grows by at least $2^{\rank(b)}$, causing its rank to change.  It follows that the total of all such charges to all nodes at all ranks is at most $2n \lg n$, $n \lg n$ for charges that do not cause rank increases and $n \lg n$ for those that do.  Suppose a node $x$ receives a residual charge, of $k$ or less.  The next merge step increases the rank of $x$ to at least $\rank(t) = \rank(y)$ (by at least $k$) because $t$ becomes a descendant of $x$.  Thus the residual charge to $x$ is at most its rank increase, and the total of all such charges is at most $n \lg n$.  \proofend

\subsection{Solid Paths as Search Trees}
\label{sec:merge-representation}

If we represent solid paths by certain kinds binary search trees, we are able to obtain a logarithmic time bound for merging.  We present three different solutions.  The first two use different kinds of finger search trees.  A \emph{finger search tree} is a form of search tree that supports an insertion or deletion at a given position in constant amortized time and a search from a given position to a position $d$ away in $O(\log(d + 2))$ time.  The type of finger search tree that applies most directly to our problem is a \emph{homogeneous finger search tree}, such as a homogeneous red-black finger search tree~\cite{finger_trees:tvw88}.  This is a red-black tree whose leaves in left-to-right order store the items of a list, in our case the nodes of a solid path in top-to-bottom order.  Every tree node contains pointers to its left and right children and to its parent.  In addition, every black node has \emph{level links} connecting it to its left and right neighbors at the same black height.  The internal nodes contain values, derived from the items, that make searching efficient.  This data structure supports insertion or deletion in constant amortized time, given a pointer to the position of the insertion or deletion.  It also supports $t = \mathit{topmost}(y, x)$ in $O(\log|P[p(t), y]| + 1)$ worst-case time.  For details see~\cite{finger_trees:tvw88}.  (But be aware that the captions are reversed on Figures 22 and 23 of that paper.)  If  we represent each solid path by a homogeneous red-black finger search tree, an $O(\log n)$ amortized time for merging follows immediately from Corollary \ref{cor:number-of-steps} and Lemma \ref{lemma:topmost}.

Homogeneous finger search trees are actually a heavyweight data structure to use in our situation.  A simpler data structure that works, although not quite as directly, is a \emph{heterogeneous finger search tree}, such as a heterogeneous red-black finger search tree~\cite{finger_trees:tvw88}.  This is a red-black tree with the items of a list stored in its nodes in symmetric order.  (In \cite{finger_trees:tvw88} the items are stored in the leaves, but it is simpler to store the items in the internal nodes, and the same time bounds hold.)  Each node contains pointers to its left and right children, except that the pointers along the left and right \emph{spines} (the paths from the root to the first and last node, respectively) are reversed: every node on the left (right) spine points to its right (left) child and to its parent.  Access to the tree is by pointers to the first and last nodes.  This data structure supports an insertion or deletion at a position $d$ away from either end in $O(\log(d + 2))$ time.  It also supports a search, such as a topmost query, to a position $d$ away from either end in $O(\log(d + 2))$ worst-case time.  Finally, it supports catenation of two trees (if the order of their items is compatible) in $O(1)$ amortized time, and splitting a tree in two at a position $d$ away from either end in $O(\log(d + 2))$ time.

To obtain an $O(\log n)$ amortized time for merging using heterogenous red-black finger search trees, we represent each solid path by a search tree, except that we split the paths containing $x$ and $y$ in two, just above $x$ and $y$, respectively.  Then all insertions and deletions are at the ends of paths, so each one takes $O(1)$ amortized time. This includes the insertions and deletions at the end of the merge, when $w$ becomes the parent of $x$. After each topmost query, we split the path containing the returned node just above that node.  When updating $x$ and $y$, we do a catenation if necessary to reflect the new state.  Each query $t = \mathit{topmost}(y, x)$ in a merge step takes $O(\log|P[p(t), y]|+1)$ time, as does the split just above $t$, because when this query is done $y$ is at one end of the path containing $t$. Thus the time for the query and the split is at most $O(1)$ plus a constant times the cost of the query. An $O(\log n)$ amortized time for merging follows from Corollary \ref{cor:number-of-steps} and Lemma \ref{lemma:topmost}.

We note that adding parent pointers for all nodes in each red-black tree, as well as doubly linking the nodes in symmetric order, allows an insertion or deletion at an arbitrary position to be done in $O(1)$ time.  In our application the solid paths are already doubly linked, but the ability to do arbitrary constant-time insertions and deletions does not help us, because this representation does not support fast searching from an arbitrary position, and splitting the paths to speed up the topmost queries results in all the insertions and deletions being at the ends of paths.  Thus this representation does not help us here. 

An even simpler data structure that (almost) works is the \emph{splay tree}~\cite{ST85}, a form of self-adjusting search tree.  This is thanks to the amazing proof by Richard Cole et al.~\cite{dynamic_finger:c00,dynamic_finger:cmss00} that splay trees are as efficient as finger search trees in the amortized sense.  Specifically, Cole's~\cite{dynamic_finger:c00} proof of the dynamic finger conjecture for splay trees gives the following bound.  Consider a splay tree representing a solid path, initially a single node, on which a sequence of insertions, deletions, and topmost queries is done.  Let $f_i$ be the \emph{finger} of the $i^{\mathrm{th}}$ operation, defined to be the node inserted in the case of an insertion, the new top node in the case of a deletion, or the node returned in the case of a topmost query.  Let $f_0$ be the single node on the initial path.  Then the amortized time of the $i^{\mathrm{th}}$ operation is $O(\log(d + 1))$, where $d$ is the number of nodes between $f_{i - 1}$ and $f_i$, inclusive, in the tree just after the operation.

To get a logarithmic bound for merging, we combine this bound with an additional amortization.  We also need to delay certain problematic insertions of nodes into the bottom of solid paths.  To do this we represent each solid path by two parts: the \emph{top part}, represented by a splay tree, and the \emph{bottom part}, which is a doubly-linked list of its nodes, top-to-bottom.  The parent and solid child pointers provide the necessary links.  We insert a node into the bottom part merely by inserting it into the doubly-linked list.  We move nodes from the bottom part into the top part only when they are involved in a topmost query.  Specifically, to do a query $t = \mathit{topmost}(y, x)$, if the bottom node of the top part is less than $x$, delete nodes one-by-one from the bottom part and insert them into the top part (the splay tree) until reaching a node greater than $x$;  return this node as $t$ (leave it in the bottom part).  Otherwise, do the query on the top part, as a splay tree operation. We call this the \emph{hybrid representation} of solid paths.

\begin{theorem}
\label{theorem:splay}
With the hybrid representation of solid paths, the amortized time per merge is $O(\log n)$.
\end{theorem}
\proof Assume that the running times are scaled so that the time bound per splay tree operation is $\lg(d + 1)$.  We shall show that the total time for all the insertions, deletions, and topmost queries is $O(1)$ per operation plus a constant times the total cost of the topmost queries.  The theorem then follows from Corollary \ref{cor:number-of-steps} and Lemma \ref{lemma:topmost}.

As merges proceed, we keep track of the locations where previous splay tree operations occurred and where future operations can occur.  We will define a potential function based on these locations, from which we derive the bound.  We define the \emph{finger} of a splay tree to be the finger of the most recent operation on it. A merge has one or two \emph{current nodes}, $x$ and possibly $y$.  Node $y$ is initially current.  It becomes non-current each time it is popped from its solid path and current each time it is updated (replaced by $t$). Given a solid path, let $d_e$ be the number of nodes less than or equal to its current node if it has one, zero if not, and let $d_f$ be the number of nodes less than or equal to its finger if it has one, zero if not (the top part is empty). We define the potential of the path to be $\lg(d_e + 1) + \lg(|d_e - d_f| + 1)$.  The total potential is the sum of the potentials of all the solid paths.

This potential has several important properties.  It is initially zero and is always non-negative, so the sum of the amortized times of a sequence of operations is an upper bound on the sum of their actual times. The following operations take $O(1)$ amortized time: creating a one-node solid path; removing the current node of a solid path ($d_e$ becomes zero, which does not increase the potential); making the top node of a solid path current ($d_e$ changes from zero to one, increasing the potential by $O(1)$);  and moving the current node to the finger (the potential does not increase).  Moving the finger to the current node decreases the potential by $\lg(|d_e - d_f| + 1)$, which makes the amortized time for an insertion or deletion at the current node, or just above or below it, $O(1)$.

Merge initialization takes $O(1)$ time and increases the potential by $O(\log n)$, for a total of $O(\log n)$ amortized time.  Consider a merge step.  Each insertion of a node into a bottom part takes $O(1)$ time.  If $x$ is in the top part of its path, each insertion above $x$ takes $O(1)$ amortized time since $x$ is a current node.  On the solid path containing $y$, a topmost query and one or more deletions at the top may be done.  Each deletion from the top takes $O(1)$ time if the top part is empty,  $O(1)$ amortized time if it is non-empty, because the first such deletion is of the current node $y$ and each subsequent deletion is at the finger.  Consider  a  query $t = \mathit{topmost}(y, x)$.  The cost of the query is $\lg|P[p(t), y]|$.  If $t$ is in the top part, the query is done as a splay tree operation and the finger moves to $f' = t$. The time for the query by Cole's bound is $\lg(|d_{f'} - d_f| + 1)$.  The amortized time for the query is
\begin{eqnarray*}
\lg(|d_{f'} - d_f| + 1) + \lg(|d_{f'} - d_e| + 1) - \lg(|d_f - d_e|) + 1)  \le  \\
2 \lg(|d_{f'} - d_e| + 1) + O(1) = 2 \lg|P[p(t), y]| + O(1).
\end{eqnarray*}
That is, the amortized time for the query is $O(1)$ plus at most twice the cost of the query.  If $t$ is the top node of the bottom part, the query takes $O(1)$ time and there are no splay tree operations.  If $t$ is in the bottom part but not the top node, each node on the bottom part above $t$ is inserted into the splay tree, and the finger moves to $f'=p(t)$, now in the top part.  If there are $k$ insertions into the top part, the time of the query by Cole's bound is at most $\lg(d_{f'} - d_f + 1)$ for the first insertion plus $O(k)$ for the rest.
The amortized time for the query is at most
$$
    \lg(d_{f'} - d_f + 1) + \lg(d_{f'} - d_e +1) - \lg(|d_f - d_e| + 1) + O(k)
$$
which is $O(k)$ plus at most twice the cost of the query.

Finally, consider the effect of updating $x$ and $y$ at the end of the merge step.  This makes $y = t$ a new current node.  If $t$ is the result of a topmost query, $t$ or its parent is the finger of its path; otherwise, $t$ is the top of its path.  In either case, making $t$ a current node increases the potential of its path by only $O(1)$.  \proofend

We conjecture that Theorem \ref{theorem:splay} is true if solid paths are represented entirely by splay trees.  We claimed such a result in the conference version of our paper~\cite{GTW06}, but our proof is incorrect.  With such a representation, the proof of Theorem \ref{theorem:splay} fails for certain insertions of nodes into the bottoms of solid paths.  We want such an insertion to take $O(1)$ amortized time, but this is not true for an insertion into a path not having a current node and whose finger is far from the bottom of the path.  Such an insertion can occur in a merge step after a step in which $x > s$, so that a topmost query does not occur.  One way to get a correct proof would be to extend Cole's proof of the dynamic finger conjecture to show that the extra time needed for $k$ arbitrary interspersed insertions at one end is $O(k)$.  We conjecture that this is true, but proving it may require delving into the details of Cole's very-complicated proof.

\subsection{Leaf Deletions}
\label{sec:deletions}

The easiest way to handle leaf deletions is just to ignore them, since deleted nodes play no role in future operations.  To reinsert a deleted node, we create a new version of it and treat it as a new node.  If there are enough deletions that the number of nodes decreases by a constant factor,  we may wish to entirely rebuild the data structure each time this happens.  This takes linear time, which is $O(1)$ per deletion.  With such rebuilding, the space used is always linear in the number of undeleted nodes, and the amortized merge time is logarithmic in the number of undeleted nodes.

\section{Implicit Mergeable Trees}
\label{sec:weak}

We now consider the special case of mergeable trees in which there are neither cuts nor parent queries.  In this case we need not store parent pointers, and indeed we do not need to explicitly maintain the trees at all.  Instead, we represent each mergeable tree by a dynamic tree of possibly different structure but \emph{equivalent} in that an nca query or a merge operation can be simulated by $O(1)$ dynamic tree operations.  This gives us an $O(\log n)$ time bound for each mergeable tree operation, worst-case, amortized, or randomized, depending on the bound of the underlying dynamic tree structure.  Since the mergeable trees are implicitly represented, we call the resulting solution \emph{implicit mergeable trees}.

In order to develop this approach, we need to introduce a little terminology.  Let $T$ a rooted tree whose nodes are selected from a totally ordered set; $T$ need not be heap-ordered.  Let $v$ and $w$ be any nodes in $T$.  We denote by $T[v, w]$ the (unique) path connecting $v$ and $w$ in $T$, ignoring arc directions.  In general this path consists of two parts, connecting $v$ and $w$, respectively, with $\nca(v, w)$.  When used in an argument of $\min$, $T$ and $T[v,w]$ denote the node sets of $T$ and $T[v, w]$, respectively.  If $T$ is heap-ordered, $\rootn(v) = \min(T)$ and $\nca(v, w) = \min(T[v, w])$.  Thus we can find roots and nearest common ancestors by computing minima over appropriate sets.  Furthermore, we need not do this the original tree; we can use any tree $T'$ that is \emph{equivalent} to $T$ in the following sense: $T$ and $T'$ have the same node sets and $\min(T[v, w]) = \min(T'[v, w])$ for all pairs of nodes $v$, $w$ in $T$.

Thus we shall represent a forest of mergeable trees by a forest of equivalent dynamic trees in which we simulate each merge by a link, or by a cut followed by a link.  We need the following additional operations on rooted but not necessarily heap-ordered dynamic trees:
\begin{itemize}
\item $\mathit{treemin}(v)$: Return the minimum node in the tree containing $v$.
\item $\mathit{pathmin}(v)$: Return the minimum node on the path from $v$ to $\rootn(v)$.
\item $\mathit{evert}(v)$: Make $v$ the root of the tree containing it, by reversing the direction of each arc on the path $P[v, \nul]$.
\end{itemize}
These dynamic tree operations are standard: see \cite{AHTdL05,GGT91,ST83,ST85,TW05}.  We implement the mergeable tree operations, excluding parent and cut, by simulating them on the equivalent dynamic trees  as follows:
\begin{itemize}
\item $\rootn(v)$: Return $\treemin(v)$.
\item $\nca(v, w)$: If $\rootn(v) \neq \rootn(w)$, return null. Otherwise, do $\mathit{evert}(v)$ and return $\pathmin(w)$.
\item $\mathit{insert}(v)$: Create a new dynamic tree having the single node $v$.
\item $\mathit{delete}(v)$: Use the method in Section \ref{sec:deletions}.  Specifically, ignore leaf deletions; optionally, rebuild the entire forest each time the number of nodes decreases by a constant factor.
\item $\merge(v, w)$: If $\rootn(v) \neq \rootn(w)$, do $\mathit{evert}(v)$ and then $\mathit{link}(v, w)$. Otherwise, do $\mathit{evert}(v)$ and let $u=\pathmin(w)$; if $u \not\in \{v, w\}$, do $\mathit{cut}(u)$ and then $\mathit{link}(v, w)$.
\end{itemize}
We shall show that the dynamic trees maintained by the implementation of the merge operations are equivalent to the corresponding mergeable trees.  Assuming that this is true, the \emph{root} and \emph{nca} functions return the correct values: if $v$ and $w$ are in the same  mergeable tree $T$, then they will be in the same dynamic tree $T'$; the value returned by $\rootn(v)$  is $\min(T') = \min(T) = \rootn_{T}(v)$, and the value returned by $\nca(v, w)$ is $\min(T'[v, w]) = \min(T[v, w]) = \nca_{T}(v, w)$, where the subscript ``$T$'' indicates the tree in which the value (root or nca) is defined.

In an operation $\merge(v, w)$, if $v$ and $w$ are in the same mergeable tree $T$ and the same dynamic tree $T'$,  $u = \nca_T(v, w)$.  If $v$ and $w$ are unrelated in $T$, the merge cuts the first arc on the path in $T'$ connecting $u$ and $v$, and then links $v$ and $w$.

It remains to show that the implementation of merging maintains equivalence.  We do this by a sequence of lemmas. We start with the simpler case, that of a merge that combines two different mergeable trees.  Suppose $v$ and $w$ are in different mergeable trees $T_1$ and $T_2$, respectively, and let $T$ be the mergeable tree produced by the operation $\merge(v, w)$.  Let $x$ and $y$ be nodes in $T$.  Assume without loss of generality (which we can do by the symmetry of $v$ and $w$ and $x$ and $y$, respectively) that $x$ is in $T_1$.

\begin{figure*}[tb]
\begin{center}
\resizebox{1.\textwidth}{!} {\includegraphics{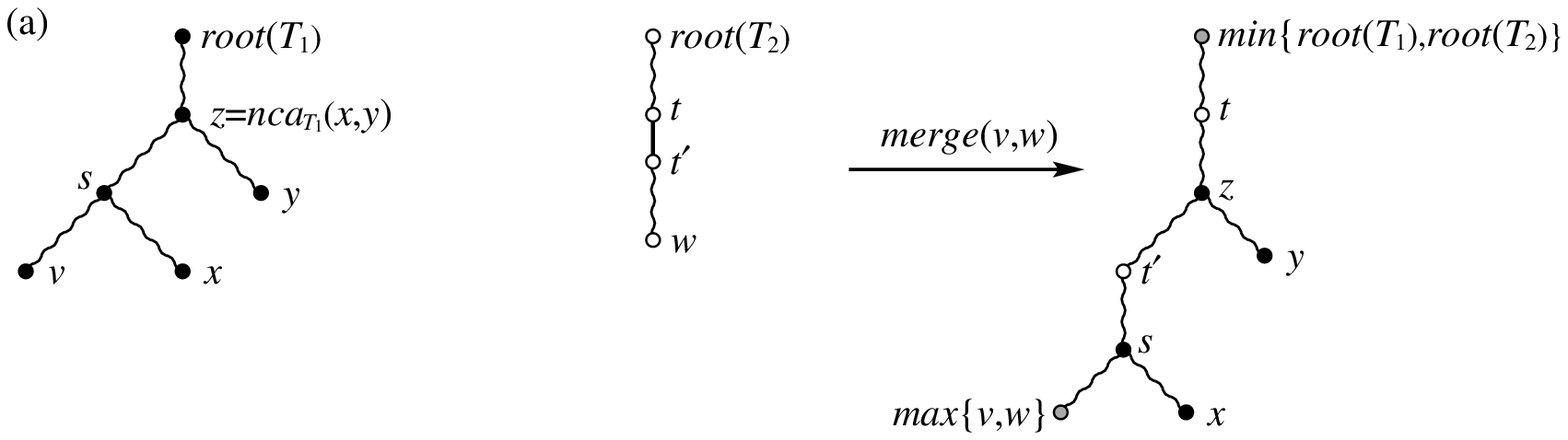}}\\
\vspace{1cm} \resizebox{1.\textwidth}{!}
{\includegraphics{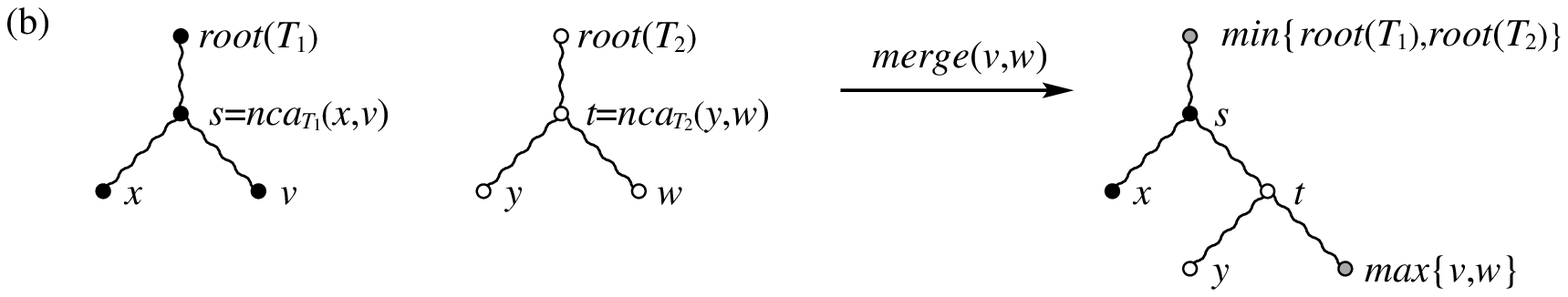}}
\end{center}
\caption{\label{fig:weak-merge-1} Proof of Lemma~\ref{lem:weak1}.  Nodes in the tree containing $v$ are black and nodes in the tree containing $w$ are white. Grey nodes can be either black or white depending on the node labels. Solid lines are single arcs; wavy lines are tree paths. (a) Node $y$ is in $T_1$. Here we assume $t<z<t'<s$.  After the merge, $\min \{v,w\}$ is on $T[t',\max\{v,w\}]$ and $\max \{\rootn(T_1),\rootn(T_2)\}$ is on $T[\min \{\rootn(T_1),\rootn(T_2)\},z]$. (b) Node $y$ is in $T_2$. Here we assume $s<t$.  After the merge, $\min \{v,w\}$ is on $T[s,\max\{v,w\}]$ and $\max \{\rootn(T_1),\rootn(T_2)\}$ is on $T[\min \{\rootn(T_1),\rootn(T_2)\},t]$.}
\end{figure*}

\begin{lemma}
\label{lem:weak1} If $y$ is in $T_1$, then $\nca_T(x, y) = \min(T_1[x,y])$.  If $y$ is in $T_2$, then $\nca_T(x, y) = \min(T_1[x, v] \cup T_2[y, w])$.
\end{lemma}
\proof
Suppose $y$ is in $T_1$. (See Figure \ref{fig:weak-merge-1}(a).) Let $z = \nca_{T_1}(x, y) = \min(T_1[x, y])$. The effect of the merge on the path between $x$ and $y$ is to insert zero or more nodes of $T_2$ into either the part of the path from $x$ to $z$ or into the part of the path from $z$ to $y$.  Any such inserted node must be larger than $z$.  Thus $\nca_{T}(x, y) = \min(T[x, y]) = z$, giving the first part of the lemma.  Suppose that $y$ is in $T_2$. (See Figure \ref{fig:weak-merge-1}(b).) Let $s = nca_{T_1}(x, v)$ and $t = \nca_{T_2}(y, w)$.  In $T$, $s$ and $t$ are related.  The path $T[x,y]$ is a catenation of $T_1[x,s]$, a path of descendants of $\min\{s, t\}$, and $T_2[t,y]$.  Thus $\nca_{T}(x, y) = \min\{s, t\} = \min(T_1[x, v] \cup T_2[w, y])$, giving the second part of the lemma. \proofend

Now suppose that $T'_1$ and $T'_2$ are trees equivalent to $T_1$ and $T_2$, respectively, and that $T'$ is formed from $T_1'$ and
$T_2'$ by rerooting $T'_1$ at $v$ and adding the arc $(v,w)$.

\begin{lemma}\label{lem:weak2} Tree $T'$ is equivalent to $T$.
\end{lemma}
\proof Clearly $T$ and $T'$ contain the same nodes.  We need to show that $\nca_{T}(x, y) = \min(T'[x, y])$ for every pair of nodes $x$, $y$, in $T$.  Assume without loss of generality that $x$ is in $T_1$.  If $y$ is in $T_1$, then $T'[x,y] = T_1'[x,y]$, and $\nca_{T}(x, y) = \min(T'[x,y])$ follows from the first part of Lemma~\ref{lem:weak1} and the equivalence of $T_1$ and $T_1'$.  If $y$ is in $T_2$, then $T'[x,y]$ is a catenation of $T'_1[x,v]$ and $T'_2[w,y]$, and $\nca_{T}(x, y) = \min(T'[x, y])$ follows from the second part of Lemma~\ref{lem:weak1} and the equivalence of $T_1$ and $T_1'$ and of $T_2$ and $T_2'$. \proofend

The case of a merge that restructures a single tree is similar but more complicated.  Consider an operation $\merge(v, w)$ of two nodes that are in the same tree $T_1$.  Let $u = \nca_{T_1}(v, w)$.  Assume that $u$ is neither $v$ nor $w$; otherwise the merge does nothing.  Let $q$ be the child of $u$ that is an ancestor of $v$, let $T_2$ be the subtree of $T_1$ with root $q$, and let $T$ be the tree produced by the merge.  Finally, let $x$ and $y$ be any nodes of $T_1$.  The next lemma is the analogue of Lemma~\ref{lem:weak1} for this case.

\begin{figure*}
\begin{center}
\resizebox{1.\textwidth}{!} {\includegraphics{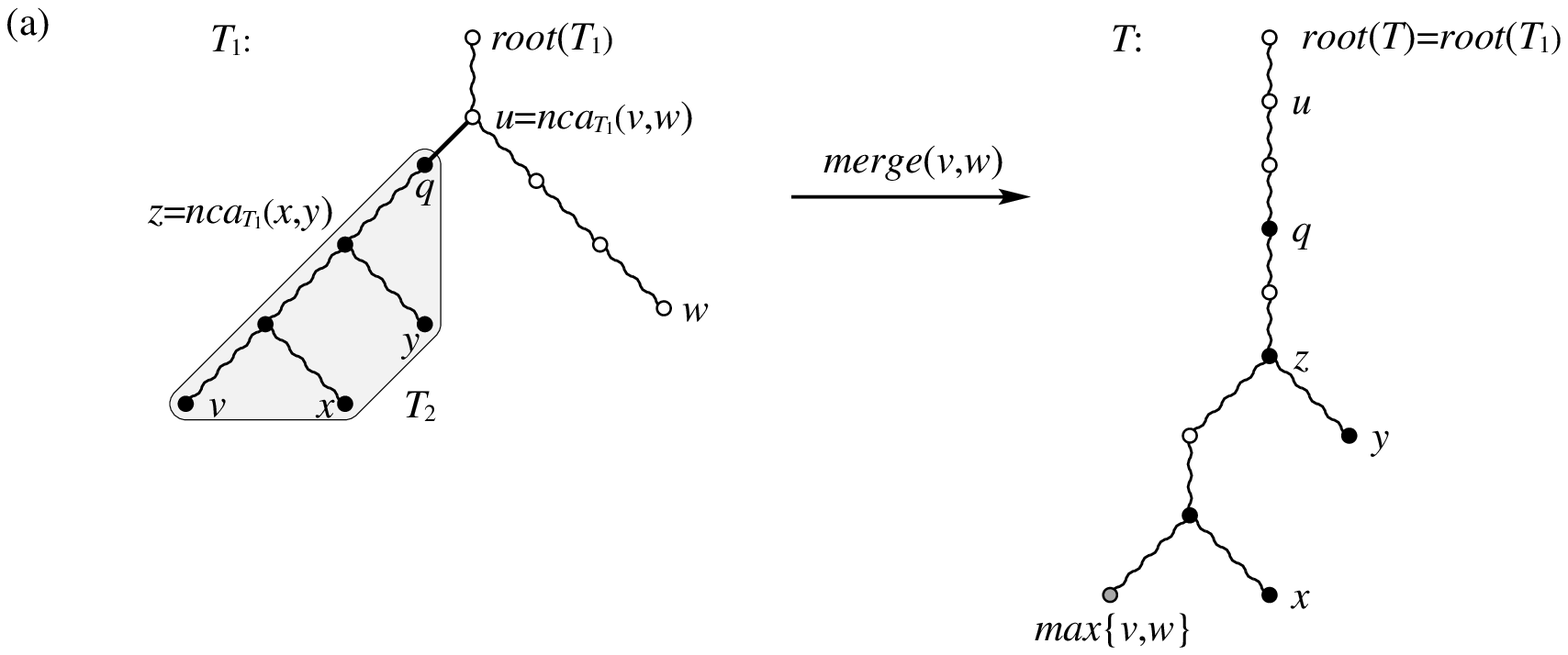}}\\
\vspace{.5cm}
\resizebox{1.\textwidth}{!} {\includegraphics{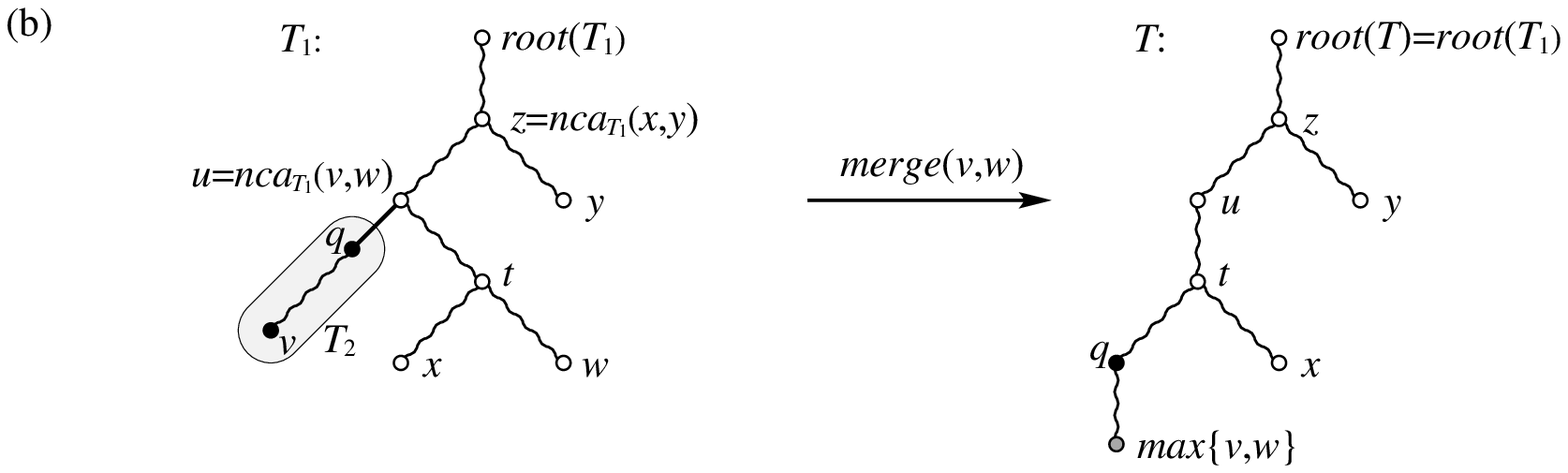}}\\
\vspace{.5cm}
\resizebox{1.\textwidth}{!} {\includegraphics{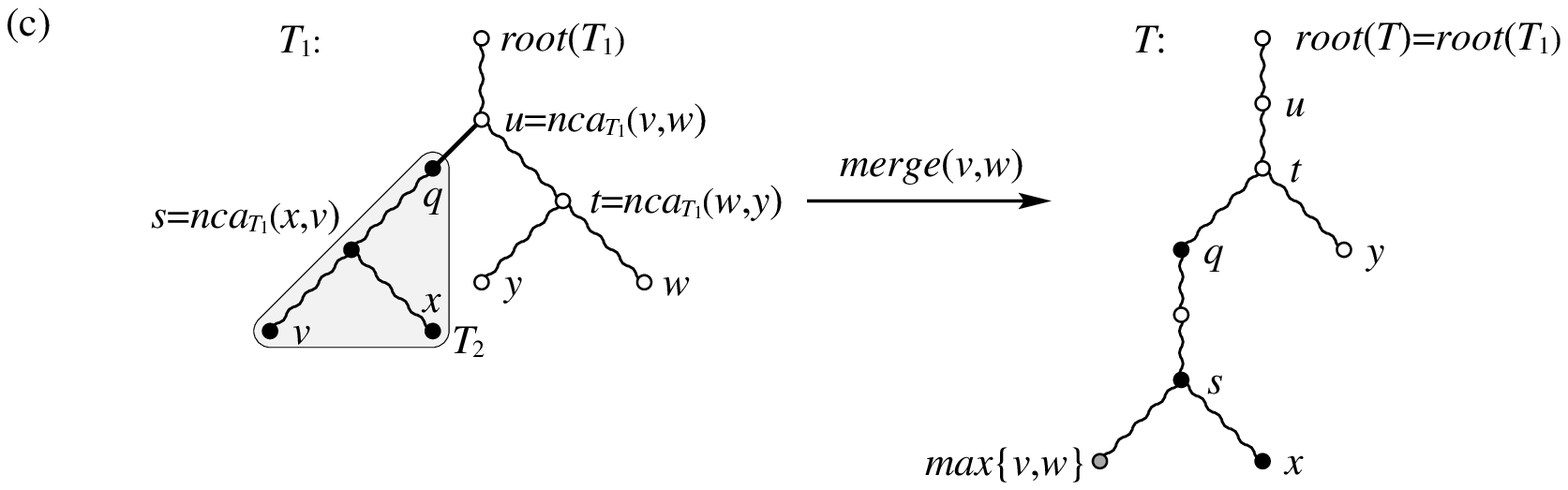}}
\end{center}
\caption{\label{fig:weak-merge-2} Proof of Lemma~\ref{lem:weak3}. Node $q$ is the child of $u$ in $T_1$ that is an ancestor of $v$.
Nodes in the subtree of $q$ are black and the rest are white. Grey nodes can be either black or white depending on the node labels.
(a) Both $x$ and $y$ are in $T_2$. After the merge, $\min \{v,w\}$ is on $T[u,\max\{v,w\}]$. (b) Neither is in $T_2$. Here we assume $t<q$.  After the merge, $\min \{v,w\}$ is on $T[t,\max\{v,w\}]$. (c) Only $x$ is in $T_2$; the situation is similar if only $y$ is in $T_2$. Here we assume $t<q$.  After the merge, $\min \{v,w\}$ is on $T[t,\max\{v,w\}]$.}
\end{figure*}

\begin{lemma}\label{lem:weak3} If both $x$ and $y$ are in $T_2$, or neither $x$ nor $y$ is in $T_2$, then $\nca_T(x, y) = \min(T_1[x,y])$.  If exactly one of $x$ and $y$, say $x$, is in $T_2$, then
$\nca_{T}(x, y) = \min(T_2[x, v] \cup T_1[w, y])$.
\end{lemma}
\proof If both $x$ and $y$ are in $T_2$, then $T_1[x,y]$ is entirely in $T_2$ (see Figure \ref{fig:weak-merge-2}(a)); if neither is in $T_2$, then $T_1[x,y]$ is entirely outside of $T_2$ (see Figure \ref{fig:weak-merge-2}(b)).  Suppose one of these cases is true.  Let $z = \nca_{T_1}(x, y) = \min(T_1[x, y])$.  The effect of the merge on the path between $x$ and $y$ is to insert into the path zero or more nodes, all of which must be larger than $z$.  Thus $\nca_{T}(x, y) = \min(T[x, y]) = z$, giving the first part of the lemma.  Suppose that exactly one of $x$ and $y$, say $x$, is in $T_2$.  Let $s = \nca_{T_1}(x, v)$ and $t = \nca_{T_1}(w, y)$. In $T$, $s$ and $t$ are related. (See Figure \ref{fig:weak-merge-2}(c)). Path $T[x,y]$ is a catenation of $T_2[x,s]$, a path of descendants of $\min\{s, t\}$ and $T_1[t,y]$. Thus $\nca_{T}(x, y) = \min(T[x, y]) = \min\{s, t\} = \min(T_2[x,v] \cup T_1[w,y])$.\proofend

Now suppose that $T'_1$ is a tree equivalent to $T_1$.  Reroot $T'_1$ at $v$, which does not affect the equivalence, and let $r$ be the parent of $u$ in $T'_1$. Deleting the arc from $u$ to $r$ breaks $T_1'$ into two trees; let $T_2'$ be the one that contains $v$ (and $r$).  Finally, let $T'$ be the tree formed from $T'_1$ by deleting the arc from $u$ to $r$ and then adding an arc from $v$ to $w$. We shall show that Lemma~\ref{lem:weak2} holds in this case; that is, $T'$ is equivalent to $T$.  This would be easy (and analogous to the proof of Lemma~\ref{lem:weak2}, but using Lemma~\ref{lem:weak3} in place of Lemma~\ref{lem:weak1}) if $T_2$ and $T_2'$ were equivalent.  This is not necessarily true, however.  Fortunately, what is true suffices for our purpose.

\begin{lemma}
\label{lem:weak4} $T_2'$ contains all the nodes in $T_2$.  Any node in $T_2'$ but not in $T_2$ is not a descendant of $u$ in $T_1$.
\end{lemma}

\begin{figure*}
\begin{center}
\resizebox{.65\textwidth}{!} {\includegraphics{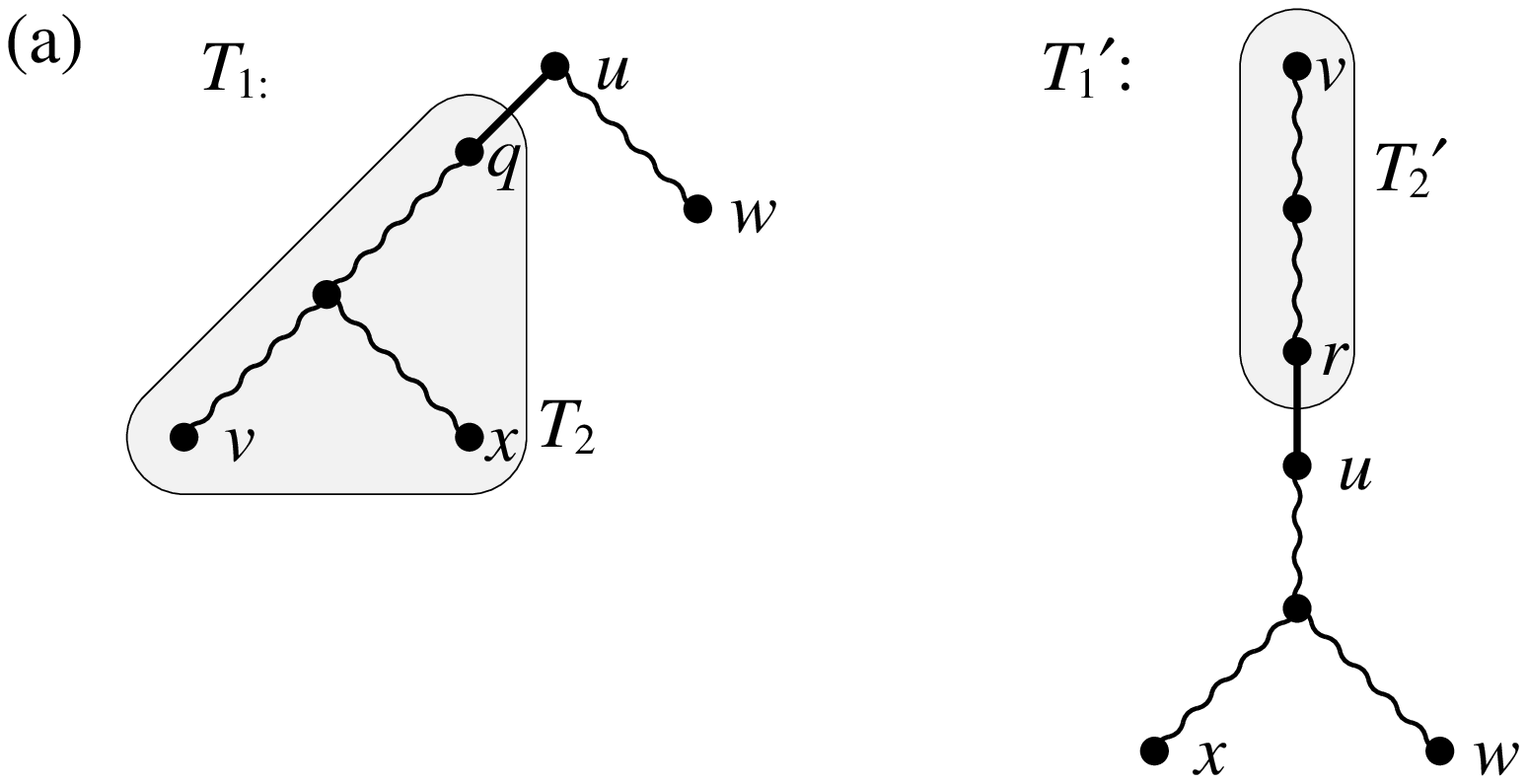}}\\
\vspace{1cm}
\resizebox{.65\textwidth}{!} {\includegraphics{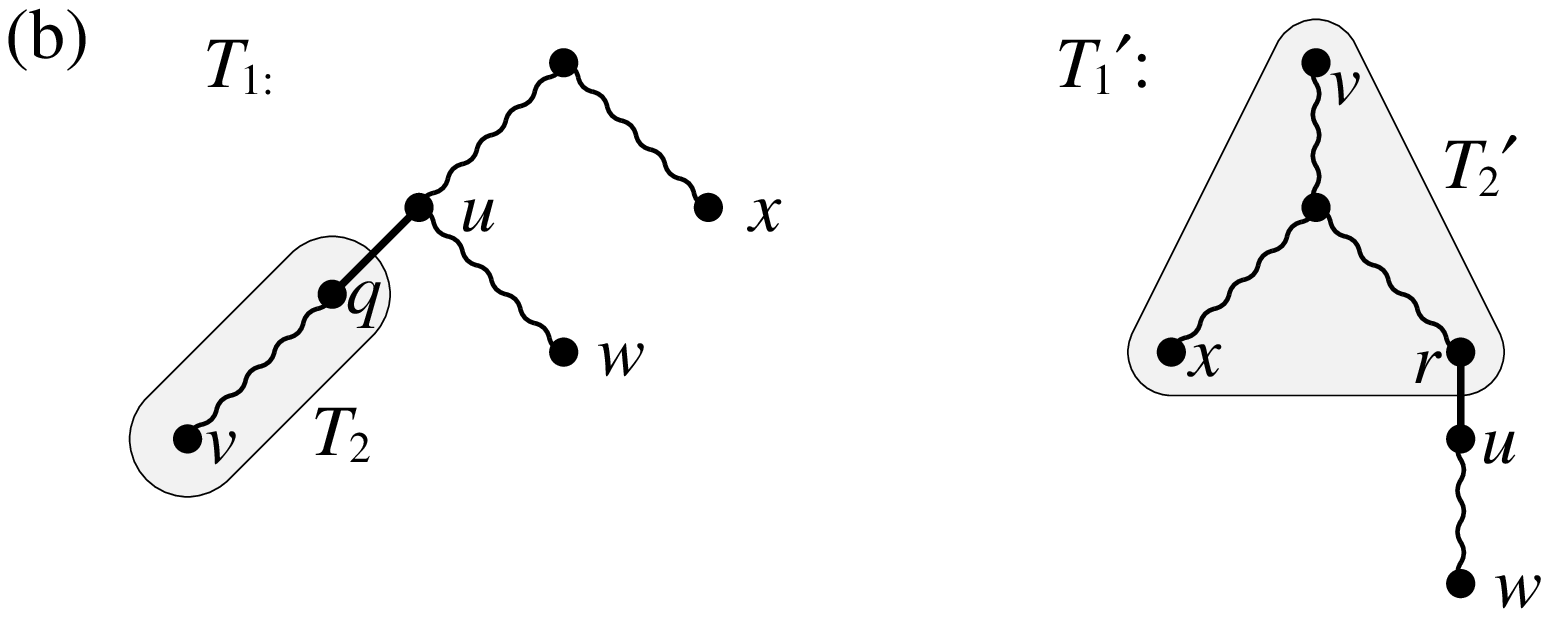}}
\end{center}
\caption{\label{fig:weak-merge-3} Proof of Lemma~\ref{lem:weak4}.
Node $u = \nca_{T_1}(v,w)$, $q$ is the child of $u$ in $T_1$ that is an ancestor of $v$, and $r$ is the parent of $u$ in $T_1'$.
(a) Assuming $x$ is not in $T'_2$ implies $\mathit{nca}_{T_1}(x,v) \le u$, a contradiction. (b) If $x$ is in $T'_2$ but not in $T_2$ then
$\nca_{T_1}(x,v) < u$.}
\end{figure*}

\proof Let $x$ be a node in $T_2$. Then $\nca_{T_1}(x, v) = \min(T_1[x, v]) \geq q > u$.  Since $T_1$ and $T_1'$ are equivalent, $\min(T'_1[x, v]) > u$.  But then $x$ must be in $T_2'$, because if it were not, $u$ would be on $T_1'[v,x]$,  which would imply
$\min(T'_1[x, v]) \leq u$, a contradiction. (See Figure \ref{fig:weak-merge-3}(a).) This gives the first part of the lemma.  Suppose $x$ is in $T_2'$ but not in $T_2$.  Since $x$ is not in $T_2$, $x$ is not a descendant of $q$ in $T_1$, which implies $\nca_{T_1}(x, v) =
\min(T_1[x, v]) \leq u$.  But since $T_1$ and $T_1'$ are equivalent, and $x$ but not $u$ is in $T_2'$, $\min(T_1[x,v]) = \min(T'_1[x, v]) = \min(T_2'[x, v]) \neq u$. (See Figure \ref{fig:weak-merge-3}(b).) Thus $\nca_{T_1}(x, v) < u$, which implies the second part of the lemma. \proofend

\begin{lemma}\label{lem:weak5} Tree $T'$ is equivalent to $T$.
\end{lemma}
\proof Trees $T$ and $T'$ contain the same nodes.  We need to show that $\nca_T(x, y) = \min(T'[x, y])$ for every pair of nodes $x$ and $y$ in $T$.  The first part of Lemma~\ref{lem:weak4} gives six cases to consider, depending upon which of the trees $T_2$ and $T_2'$ contain $x$ and $y$.  If $x$ and $y$ are both in $T_2$, or both in $T_2'$ but not in $T_2$, or both not in $T_2'$, then $\nca_T(x, y) =
\min(T_1[x, y]) = \min(T'_1[x, y]) = \min(T'[x, y])$ by the first part of Lemma~\ref{lem:weak3}, the equivalence of $T_1$ and $T_1'$, and the construction of $T'$.  If one of $x$ and $y$, say $x$, is in $T_2$, and the other, $y$, is not in $T_2'$, then $\nca_T(x,y) = \min(T_2[x, v] \cup T_1[w,y])= \min(T_2'[x, v] \cup T_1'[w, y]) = \min(T'[x, y])$ by the second part of Lemma~\ref{lem:weak3}, the equivalence of $T_1$ and $T_1'$, and the construction of $T'$.  These cases are analogous to the two cases in the proof of Lemma~\ref{lem:weak2}.

\begin{figure*}
\begin{center}
\resizebox{.65\textwidth}{!} {\includegraphics{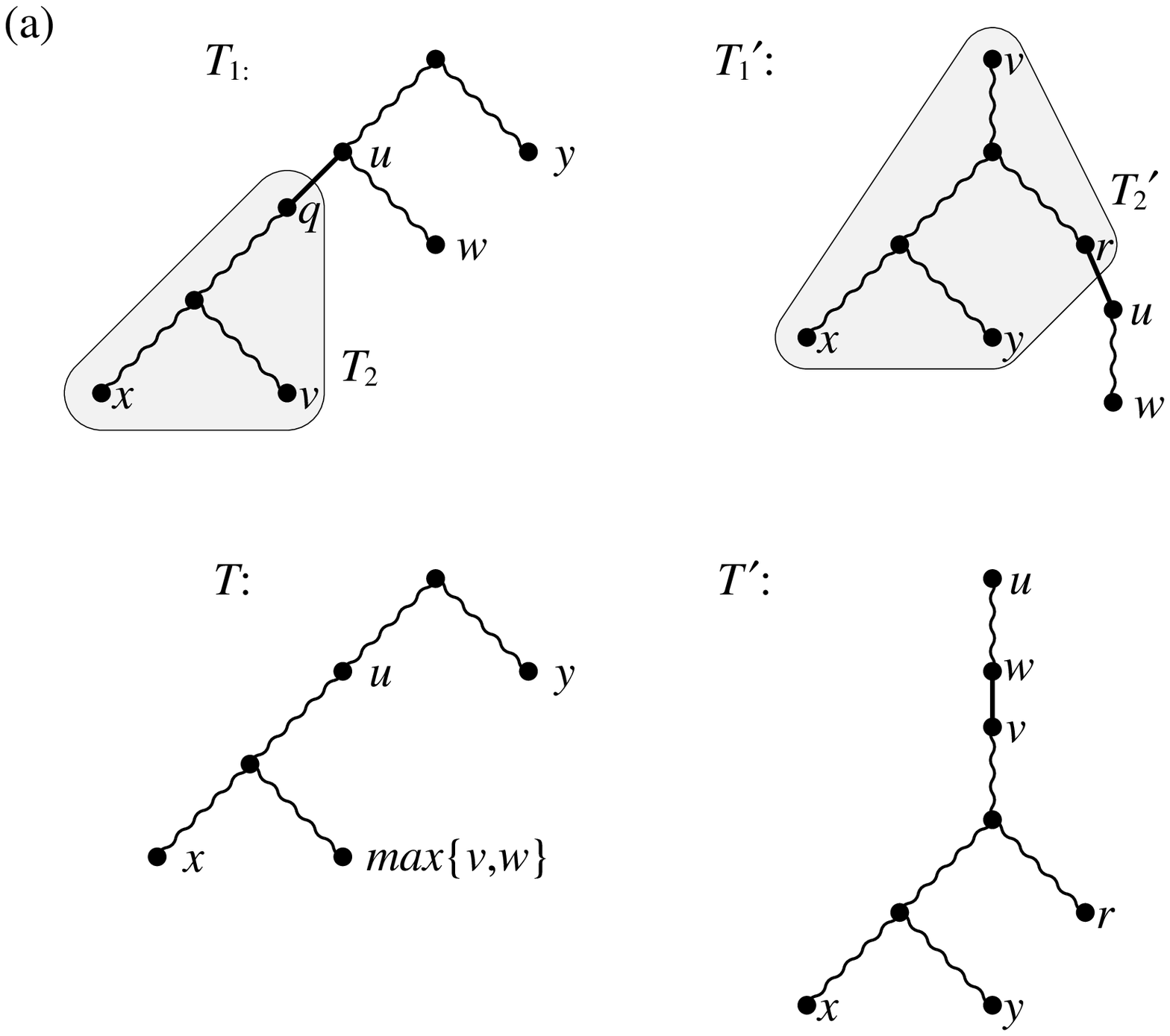}}\\
\vspace{.5cm}
\resizebox{.65\textwidth}{!} {\includegraphics{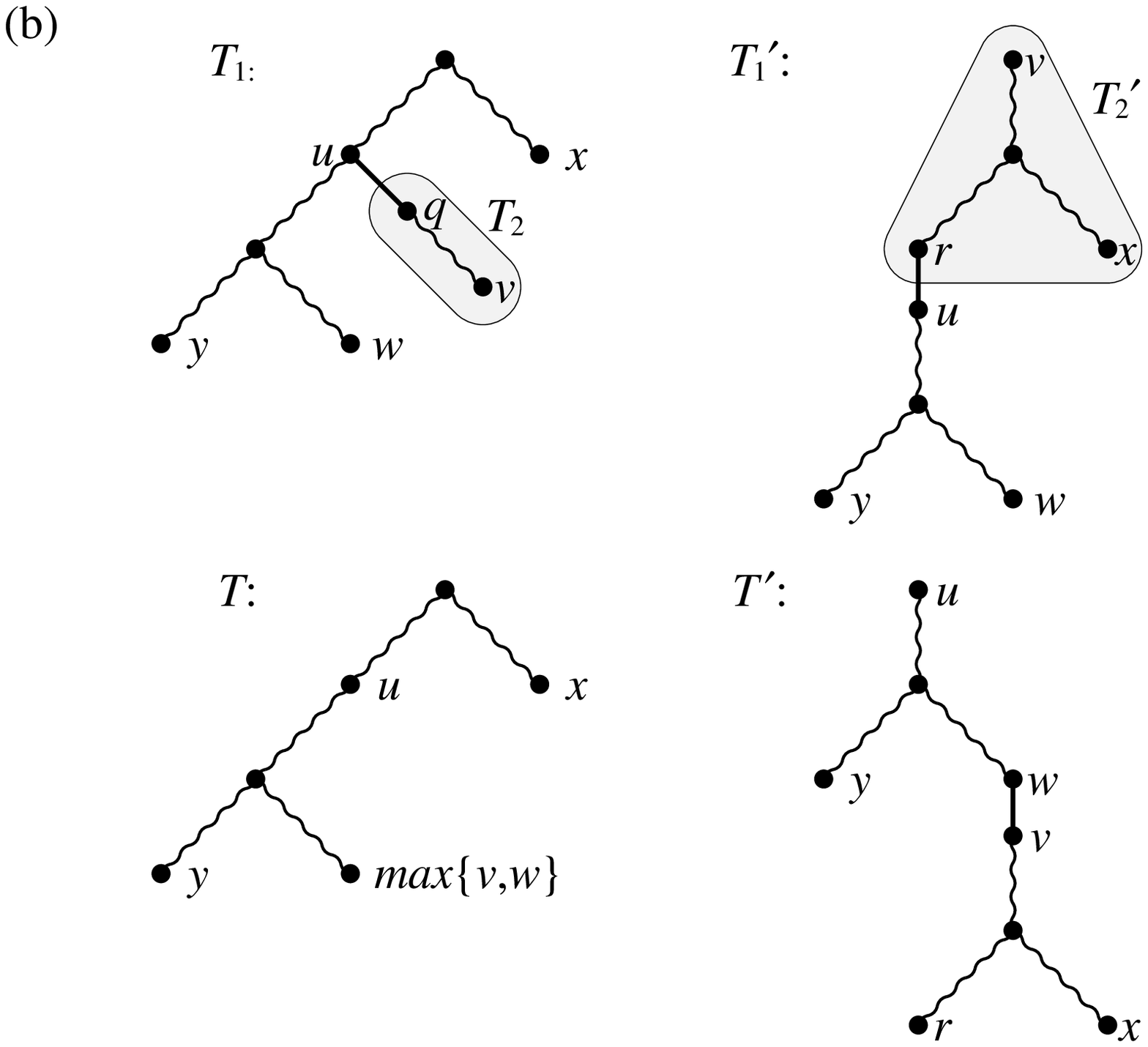}}
\end{center}
\caption{\label{fig:weak-merge-4} Proof of Lemma~\ref{lem:weak5}. Node $u = \nca_{T_1}(v,w)$, $q$ is the child of $u$ in $T_1$ that is an ancestor of $v$, and $r$ is the parent of $u$ in $T_1'$. After the merge, $\min \{v,w\}$ and $q$ are on $T[u,\max\{v,w\}]$.
(a) Node $x$ is in $T_2$ and $y$ is in $T'_2$ but not in $T_2$. Then $y$ is not a descendant of $u$ in $T_1$, so $\nca_{T_1}(x,y) < u$. Since $x \in T_2$, $\min(T_2[x,v])>u$. These two facts imply $\nca_T(x,y)=\min(T'[x,y])$. (b) Node $x$ is in $T'_2$ but not in $T_2$ and $y$ is not in $T'_2$. Then $\nca_{T}(x,y) = \min(T'_1[x,y]) < u$. All nodes in $T'_1[v,w]$ are greater than or equal to $u$; thus $\nca_T(x,y) = \min(T'[x,y])$.}
\end{figure*}

The remaining two cases are new.  Suppose one of $x$ and $y$, say $x$, is in $T_2$, and the other, $y$, is in $T_2'$ but not in $T_2$.
(See Figure \ref{fig:weak-merge-4}(a).) By the second part of Lemma~\ref{lem:weak4}, $y$ is not a descendant of $u$ in $T_1$.  Then $\nca_{T_1}(x, y) = \nca_{T_1}(w, y) = \min(T_1[w, y]) < u$. Also $\min(T_2[x, v]) > u$, since $x$ is in $T_2$.  Thus $\nca_{T}(x, y) = \min(T_1[w, y]) = \nca_{T_1}(x, y) = \min(T_1[x, y]) = \min(T'_1[x, y]) = \min(T'_2[x, y]) = \min(T'[x, y])$ by the second part of
Lemma~\ref{lem:weak3}, the equivalence of $T_1$ and $T_1'$, and the construction of $T'$.  Last, suppose one of $x$ and $y$, say $x$,
is in $T_2'$ but not in $T_2$, and the other, $y$, is not in $T_2'$. (See Figure \ref{fig:weak-merge-4}(b).) Then $u$ is on $T_1'[x,y]$.
It follows that $\nca_T(x, y) = \min(T_1[x, y]) = \min(T_1'[x, y]) \leq u$ by the first part of Lemma~\ref{lem:weak3} and the equivalence of $T_1$ and $T_1'$.  But $\min(T_1'[x, y]) = \nca_{T_1}(x, y) \neq u$ since $x$ is not a descendant of $u$ by Lemma~\ref{lem:weak4}.  Thus $\nca_T(x, y) = \min(T_1'[x, y]) < u$.  Paths $T'_1[x,y]$ and $T'[x,y]$ contain the same nodes except possibly for nodes of the path $T_1'[v,w]$, all of which must must be at least $u$, since $\min(T_1'[v, w]) = \min(T_1[v, w]) = u$. Thus $\nca_T(x, y) = \min(T_1'[x, y]) = \min(T'[x, y])$.
\proofend

Lemmas \ref{lem:weak1} and \ref{lem:weak4} give us the following theorem:
\begin{theorem}
\label{theorem:implicit} The implementation of implicit mergeable trees using dynamic trees is correct.
\end{theorem}
Thus if there are no cuts and no parent queries, we can simulate each mergeable tree operation by $O(1)$ dynamic tree operations, giving an $O(\log n)$ time bound per operation, worst-case, amortized, or randomized depending upon the efficiency of the underlying dynamic trees.  Since the roots of the dynamic trees are irrelevant to the representation, we can use unrooted dynamic trees, such as top trees~\cite{AHTdL05} or topology trees~\cite{Fre97a} instead.  We can avoid the need for the \emph{treemin} operation on dynamic trees by using a separate disjoint set data structure~\cite{Tar75,Smi90} to handle root queries.  The disjoint sets are the node sets of the trees; each root query is a find query in the disjoint set structure, and each merge of two different trees requires a union of their node sets.  The extra time per merge is $O(1)$ worst-case and the time bound per root query is logarithmic worst-case and inverse-Ackermann amortized.

It seems hard if not impossible to extend the method presented here to handle parent queries or cuts, because the connection between mergeable trees and the equivalent dynamic trees can be quite loose.  In particular, a mergeable tree that is a single path can be represented by a dynamic tree consisting of one node that is adjacent to all the other nodes in the tree: consider the sequence $\mathit{merge}(1, n)$, $\mathit{merge}(2, n)$, $\mathit{merge}(3, n)$, $\ldots$, $\mathit{merge}(n-1, n)$ applied to an initial set of singleton trees.  For such an example, performing a \emph{parent} query or a \emph{cut} on the mergeable tree will take at least $\Omega(n)$ time on the equivalent dynamic tree.

\section{Persistence Pairings via Mergeable Trees}
\label{sec:pairing}

Our motivating application for mergeable trees is a problem in computational topology, that of computing an \emph{extended persistence pairing} of the critical points of a 2-dimensional connected manifold embedded in ${\cal R}^3$.  An algorithm for this problem was proposed by Agarwal et al.~\cite{AEHW04,AEHW06}.  The use of mergeable trees in this algorithm gives an $O(n \log n)$-time implementation, where $n$ is the number of critical points.  We shall describe the pairing algorithm in some detail, because the exact form it takes affects the set of operations needed on the mergeable trees.  In particular by modifying their algorithm, we are able to avoid the need for parent queries, thereby allowing the use of the implicit mergeable trees of Section \ref{sec:weak}.  We also fill in a lacuna in their algorithm.

The critical points of a manifold are the local minima, local maxima, and saddle points in a particular direction, say increasing $z$-coordinate.  The algorithm of Agarwal et al.\ computes a directed acyclic graph called the \emph{Reeb graph} that represents the skeleton of the manifold, during a $z$-increasing sweep over the manifold.  The Reeb graph is actually a multigraph; that is, multiple arcs (arcs with the same start and end vertices) can occur. The vertices of the Reeb graph correspond to the critical points. Agarwal et al.\ assume that the manifold is perturbed so that the critical points all have different $z$-coordinates, and so that the skeleton of the manifold has no vertex of degree exceeding three. In particular, each vertex is of one of four kinds: a \emph{source}, with in-degree zero and out-degree one; a \emph{sink}, with in-degree one and out-degree zero; an \emph{up-fork}, with in-degree one and out-degree two; or a \emph{down-fork}, with in-degree two and out-degree one.  The vertices of the Reeb graph are topologically ordered by the $z$-coordinate of the corresponding critical point. We call this the \emph{canonical order} (there may be other topological orderings).

The algorithm of Agarwal et al.\ pairs the vertices of the Reeb graph, and hence the critical points of the manifold, during a sweep of the graph that visits the vertices in canonical order, modifying the graph as it proceeds.  This is the part of the algorithm that uses mergeable trees.  The pairing can be done during the sweep over the manifold that builds the graph, but for our purposes it is better to think of the pairing process as a separate sweep.  We identify each vertex with its number in canonical order.  The pairing sweep maintains three invariants: (1) each vertex, once visited, has in-degree at most one; (2) a visited vertex is paired if and only if both its in-degree and its out-degree are one, or both its in-degree and out-degree are zero; and (3) the vertex order is topological.  When visiting a vertex $x$, the pairing sweep applies the appropriate one of the following cases:
\begin{itemize}
\item Case 1: $x$ is a source. Do nothing.
\item Case 2: $x$ is an up-fork. Do nothing.
\item Case 3: $x$ is a down-fork, with incoming arcs from $v$ and $w$ (which may be equal).  Concurrently walk backward along the paths ending at $v$ and at $w$, each time taking a step back from the larger of the two vertices on the two paths, until reaching a vertex $y$ common to the two paths, or trying to take a step back from a source $y$.  Pair $x$ with $y$.  Merge the two paths traversed, arranging the vertices in order along the merged path.
\item Case 4: $x$ is a sink, with an incoming arc from $v$.  Delete $x$.  While $v$ is paired, delete $v$ and replace it by its predecessor (before the deletion).  Pair $x$ with $v$.
\end{itemize}

It is straightforward to prove by induction on the number of visited vertices that the pairing sweep maintains invariants (1)--(3).  If the manifold is connected, as we have assumed, the second alternative in invariant (2) applies only after the last vertex (a sink) is processed; it is paired (in Case 4) with the first vertex, which becomes the only vertex of in-degree and out-degree zero.  If the manifold is disconnected, there will eventually be one instance of the second alternative in invariant (2) for each connected component of the manifold, corresponding to the pairing of its global minimum and global maximum points.

The pairing sweep algorithm can be implemented directly.  Vertices need not be deleted in Case 4 but merely traversed, since each such vertex is traversed only once; deleting them merely makes the statement of invariant (2) simpler.  The running time is $O(n)$ plus the time spent walking backward along paths in Case 3, which can be $\Theta(n^2)$.  To reduce this time we use mergeable trees.

\begin{figure*}
\begin{center}
\resizebox{.22\textwidth}{!} {\includegraphics{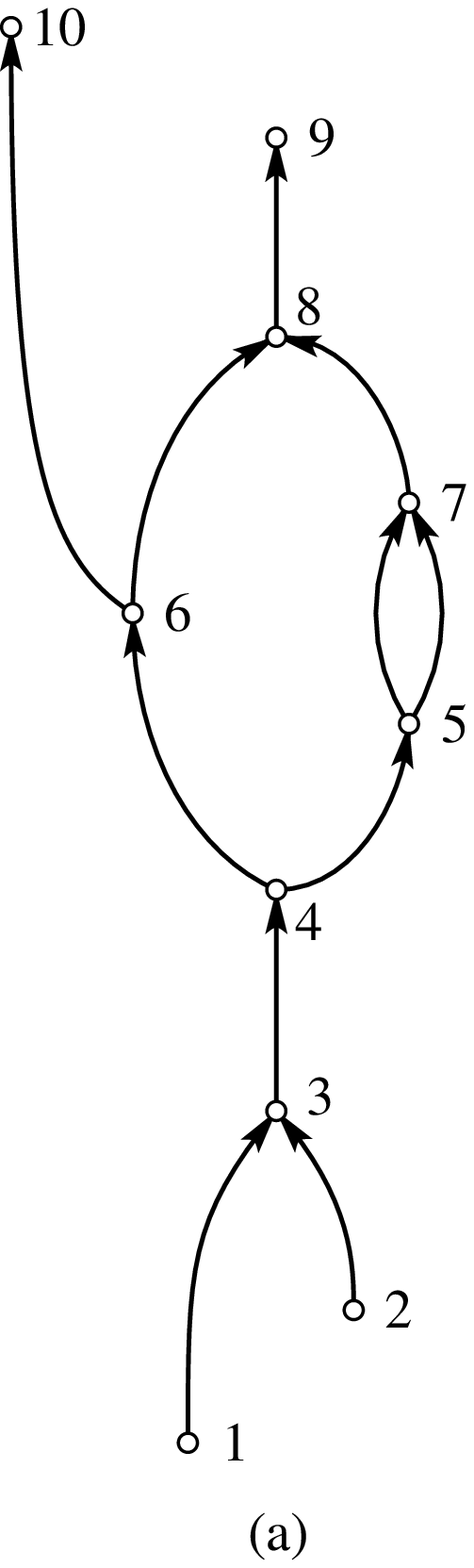}} \hspace{.3cm}
\resizebox{.22\textwidth}{!} {\includegraphics{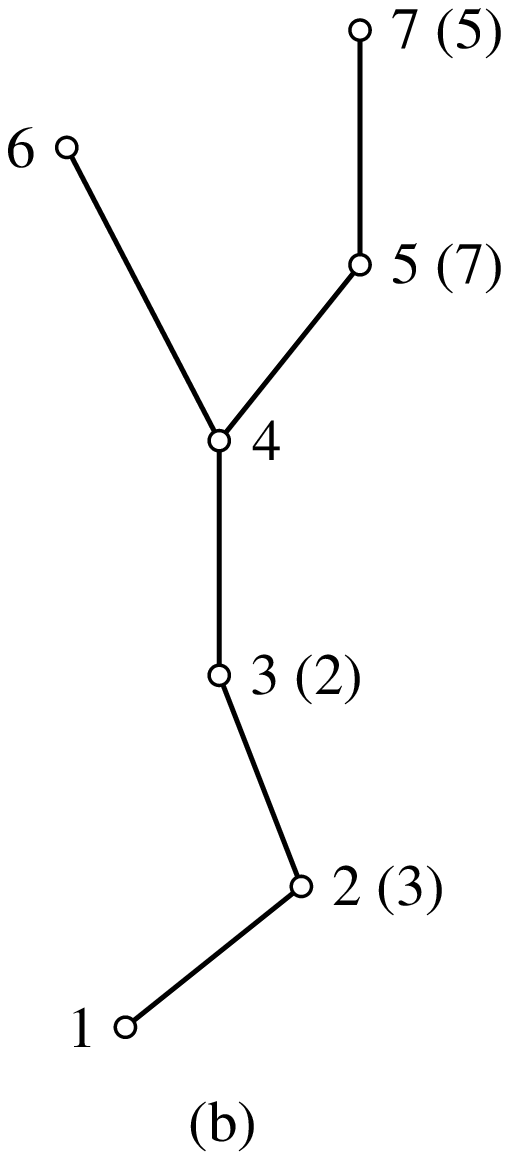}} \hspace{.3cm}
\resizebox{.22\textwidth}{!} {\includegraphics{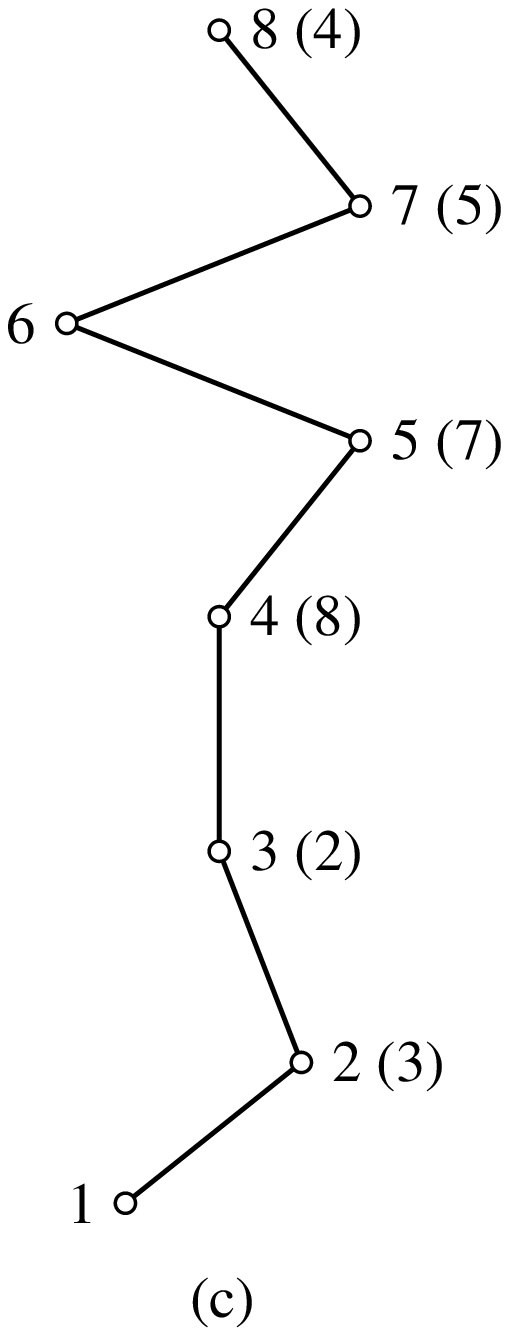}} \hspace{.3cm}
\resizebox{.22\textwidth}{!} {\includegraphics{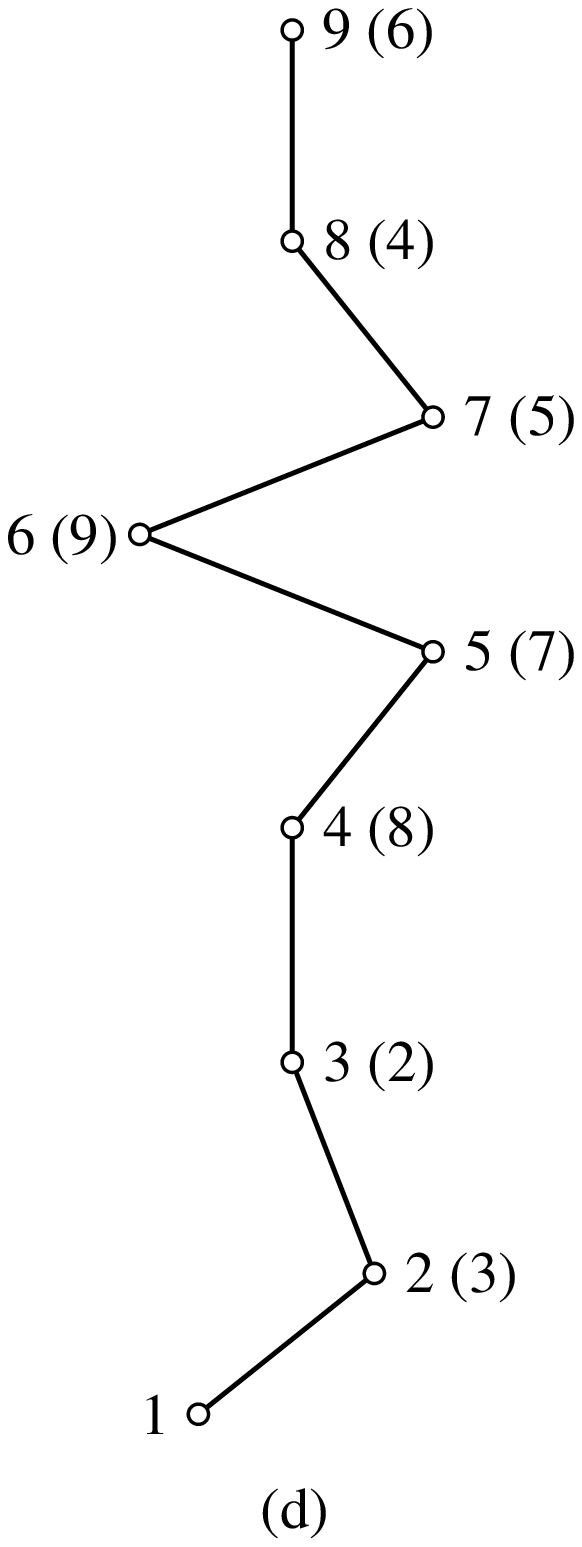}}
\end{center}
\caption{\label{fig:reeb-fwd}Example of the execution of the single-pass pairing algorithm that uses mergeable trees that support the parent operation. (a) The forward Reeb graph. (b)-(d) The mergeable tree rooted at 1.
(b) The mergeable tree after processing vertices 1 to 7. For each vertex $v$ the number in the parenthesis is the vertex paired with $v$. (c) After processing vertex 8, which is a down-fork with incoming arcs from 6 and 7. Vertex 8 is paired with $\nca(6,7)=4$ and then we perform $\merge(8,7)$ and $\merge(8,6)$. (d) After processing vertex 9; the arc (9,8) is inserted as a result of $\merge(9,8)$. To pair 9 we perform successive \emph{parent} operations, starting from 9, until we reach the first unpaired vertex, which is 6.}
\end{figure*}

Specifically, we store the set of visited vertices as the nodes of a collection of mergeable trees and perform appropriate mergeable tree operations in Cases 1--4.  When visiting a vertex $v$, we first make it into a new, one-node mergeable tree and then apply the appropriate one of the following cases:

\begin{itemize}
\item Case $1'$: $x$ is a source.  Do nothing.
\item Case $2'$: $x$ is an up-fork, with an incoming arc from $v$. Do $\mathit{merge}(x, v)$.
\item Case $3'$: $x$ is a down-fork, with incoming arcs from $v$ and $w$.  If $v$ and $w$ are in different trees, pair $x$ with $\max\{\mathit{root}(v), \mathit{root}(w)\}$; otherwise, pair $x$ with $\mathit{nca}(v, w)$.  In either case do $\mathit{merge}(x, v)$ and $\mathit{merge}(x, w)$.
\item Case $4'$: $x$ is a sink, with an incoming arc from $v$.  Do $\mathit{merge}(x, v)$.  While $v$ is paired, replace $v$ by its parent in its mergeable tree.  Pair $x$ with $v$.
\end{itemize}
See Figure~\ref{fig:reeb-fwd}. This is a restatement of the algorithm of Agarwal et al.\ that explicitly uses mergeable trees, with a lacuna corrected in Case 4: Agarwal et al.\ imply that the predecessor of $x$ is unpaired, but this need not be true.  Edelsbrunner (private communication, 2006) suggested fixing this problem by eliminating paired nodes from the mergeable trees, replacing each paired degree-two node by an arc from its child to its parent.  But we prefer the method above, since it requires no restructuring of the trees, and it leads to the two-pass pairing algorithm we develop below.  Agarwal also pair the first and last vertex separately, but this is redundant, since this pair is found by the method above in Case $4'$.

The total number of mergeable tree operations done by this method is $O(n)$, since each case except $4'$ does $O(1)$ tree operations, and the total number of tree operations done by all executions of Case $4'$ is $O(n)$: any particular vertex $v$ can be replaced by its parent in at most one execution of Case $4'$,  since such a replacement corresponds to the deletion of $v$ in the corresponding execution of Case 4.  The time spent in addition to mergeable tree operations is $O(n)$. The mergeable tree implementations discussed in Sections~\ref{sec:dyntrees} and \ref{sec:part-rank} can be used, since no cuts are needed.  The fastest method of Section~\ref{sec:part-rank} has an amortized $O(\log n)$ time bound per mergeable tree operation, giving an $O(n\log n)$ time bound for pairing.

\begin{figure*}
\begin{center}
\resizebox{.22\textwidth}{!} {\includegraphics{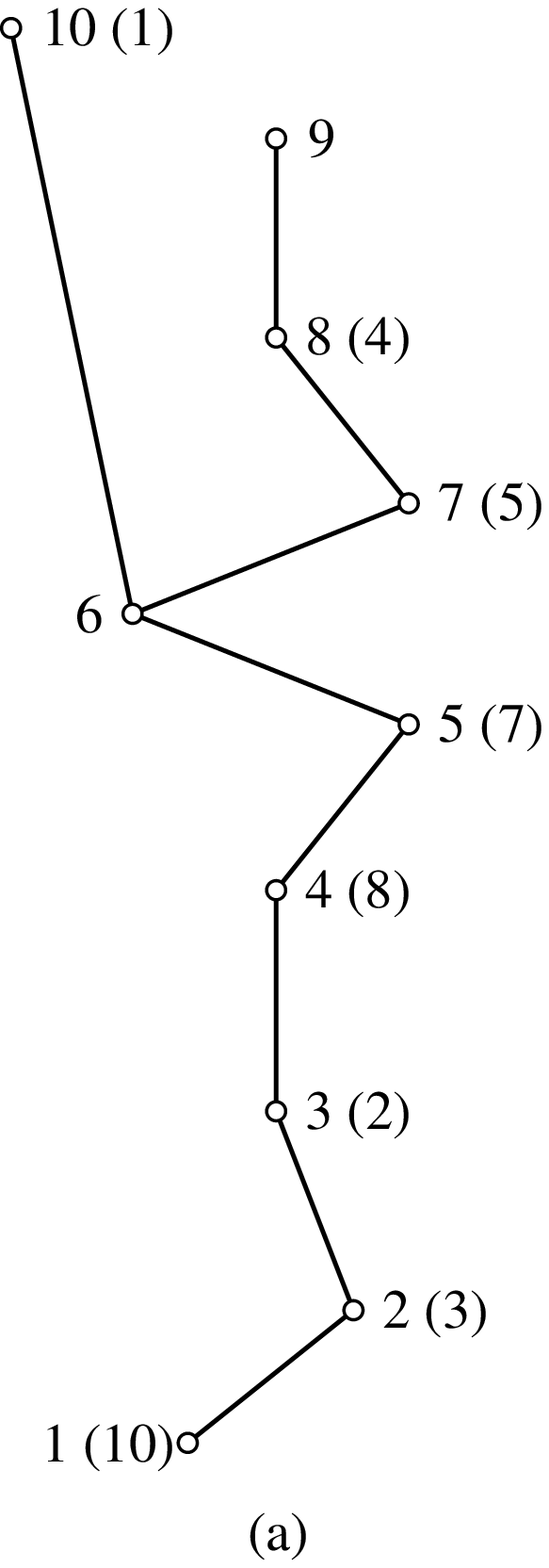}} \hspace{1.5cm}
\resizebox{.22\textwidth}{!} {\includegraphics{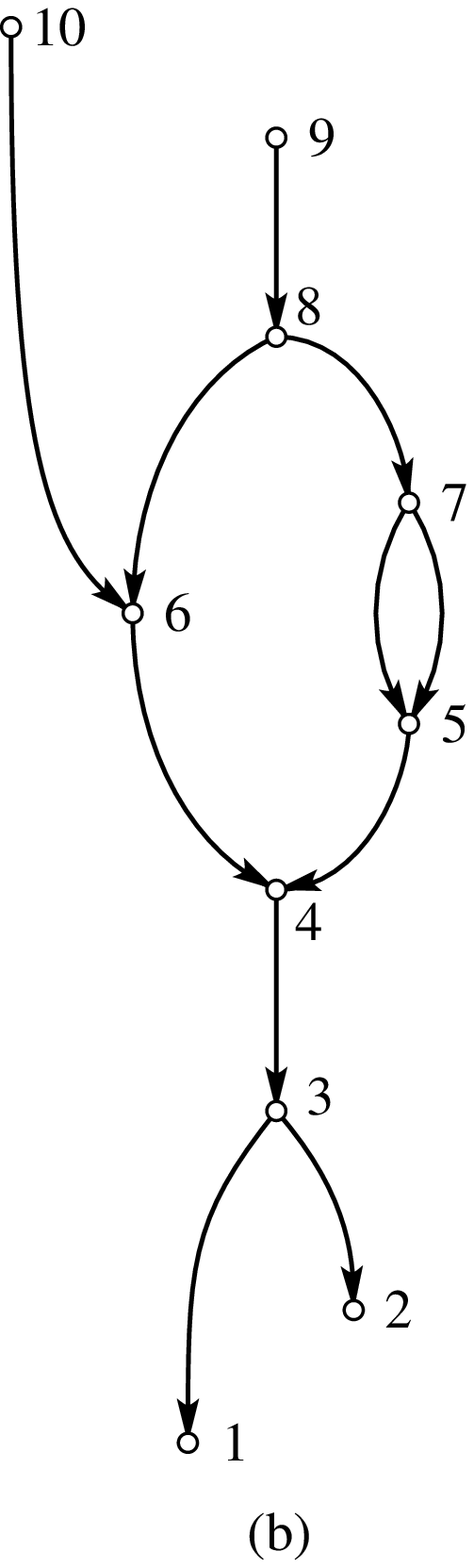}} \hspace{1.5cm}
\resizebox{.22\textwidth}{!} {\includegraphics{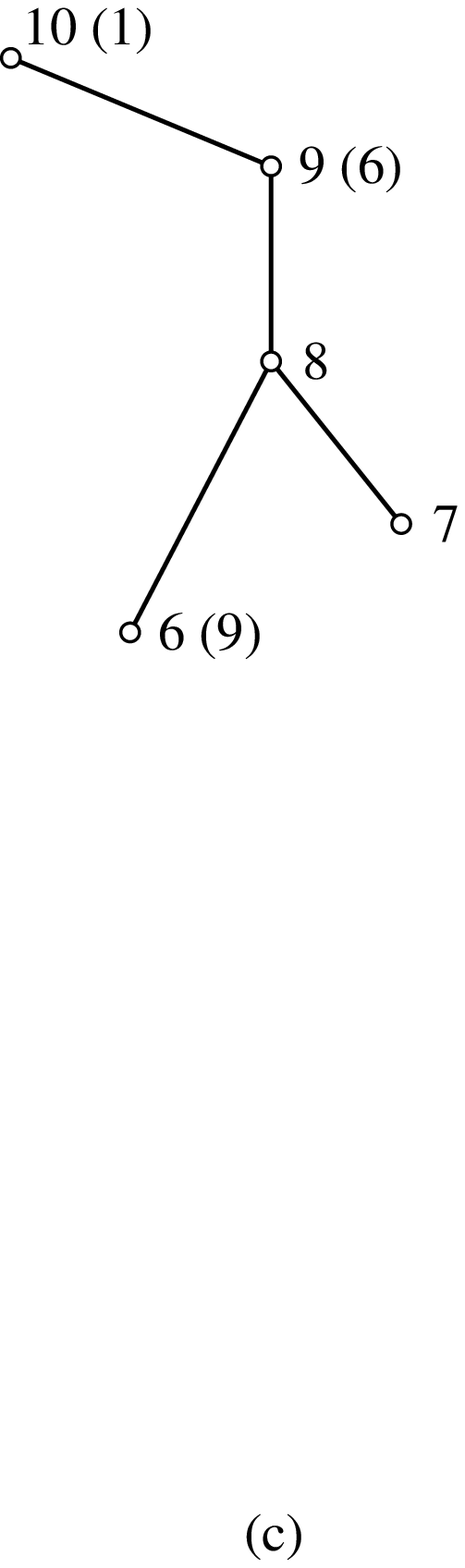}}
\end{center}
\caption{\label{fig:reeb-rev}Example of the execution of the two-pass pairing algorithm. (a) The mergeable tree (rooted at 1) produced during the first pass; all pairs except (6,9) were found. (b) The reverse Reeb graph used in the second pass. (c) The mergeable tree rooted at 10, after processing vertex 6 which is a down-fork in the reverse graph with incoming arcs from 10 and 8. Vertex 6 is paired with $\min \{ 8,10 \}=8$ (which has label greater than the label of 10 in the reverse graph) and then we perform $\merge(6,8)$ and $\merge(6,10)$.}
\end{figure*}

We can avoid the need for parent queries in the mergeable trees by doing two passes of a streamlined version of the method above, one in topological order and the other in reverse topological order.  This allows the use of the mergeable tree implementation described in Section~\ref{sec:weak}, which uses ordinary dynamic trees as a black box.  In order to obtain this result, we need an observation about the pairing that the algorithm produces.  Each pair is of one of four types: (a) a down-fork and an up-fork, found in Case 3; (b) a down-fork and a source, also found in Case 3; (c) a sink and an up-fork, found in Case 4; or (d) a source and a sink, also found in Case 4.  As mentioned above, there is exactly one source-sink pair if the manifold is connected, as we are assuming. If we reverse the direction of all the arcs of the Reeb graph, then every source becomes a sink and vice-versa, and every up-fork becomes a down-fork and vice versa.  If we run the pairing algorithm on the reversed graph using as the topological order the reverse of the original topological order, we compute the same pairing, except that every type-(b) pair becomes a type-(c) pair and vice-versa; type-(a) pairs and type-(d) pairs remain type (a) or (d), respectively.  But this means that \emph{every} pair except the unique type-(d) pair will be found in Case 3 of either the forward sweep or the reverse sweep.  Thus we can find all the pairs by pairing the first and last vertices, and running forward and reverse sweeps of the above method with Case $4'$ replaced by the following:
\begin{itemize}
\item Case $4''$: $x$ is a sink, with an incoming arc from $v$. Do $\mathit{merge}(x, v)$.
\end{itemize}
See Figure~\ref{fig:reeb-rev}. With this method the only operations needed on the mergeable trees are insert, root, nca, and merge.  Use of the mergeable tree implementation of Section~\ref{sec:weak} gives an $O(n\log n)$-time pairing algorithm.  Though this does not improve the asymptotic time bound, it avoids the complexities of Section~\ref{sec:part-rank}, and it avoids the iteration in Case 4.

\section{Complexity}
\label{sec:complexity}

\begin{figure*}
\begin{center}
\resizebox{1\textwidth}{!} {\includegraphics{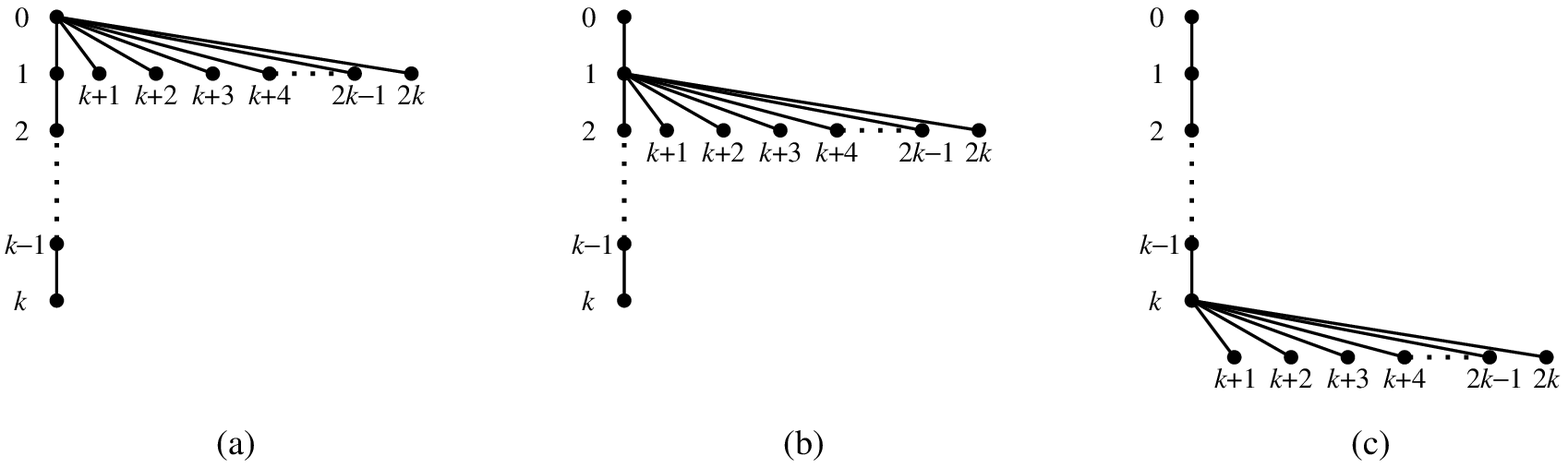}}
\end{center}
\caption{\label{fig:n2} An example with $\Theta(n^2)$ merges. The
sequence of merges consists of $k$ rounds, each with $k$ merges;
the $i^{\mathrm{th}}$ merge of round $j$ merges nodes $k+i$ and $j$. The
number of nodes is $n = 2k+1$. Figure \ref{fig:n2}(a) is the initial tree,
Figure \ref{fig:n2}(b) is the tree after the first round of $k$ merges, and
Figure \ref{fig:n2}(c) is the tree after all $k^2$ merges.}
\end{figure*}

In this section we make several observations related to the inherent complexity of the mergeable trees problem, in an effort to clarify under what circumstances further improvements or alternative methods might be possible.  We begin by bounding the number of possible merge operations in the absence of cuts.  If each merge is of two leaves, then the number of merges is at most $n - 1$, since each merge reduces the number of leaves by one; a leaf deletion cannot increase the number of leaves.  On the other hand, if merges are of arbitrary nodes, there can be $\Theta(n^2)$ merges, each of which changes the forest.  Figure \ref{fig:n2} gives an example with $\Omega(n^2)$ merges.  Since any merge that changes the forest must make at least one pair of nodes related, there can be at most $n \choose 2$ merges.

We can also bound the number of parent changes in the absence of cuts.  Section \ref{sec:dyntrees} gives a bound of $O(m + n\log n)$.  The example in Figure \ref{fig:n2} gives a lower bound of $\Omega(m)$.  The following example gives a bound of $\Omega(n \log n)$.  Combining this example and the one in Figure \ref{fig:n2} gives an example with a bound of $\Omega(m + n \log n)$, thus showing that the $O(m + n \log n)$ bound is tight. Start with $n = 2^k$ one-node trees.  Merge these in pairs to form two-node paths, then merge the pairs in pairs to form four-node paths, and so on, until there is only a single tree, consisting of a single path.  Order the nodes so that in each merge the nodes of the two merged paths are perfectly interleaved.  Then the number of merges is $n - 1$ and the number of parent changes is $n/2 +3n/4 + 7n/8 + ... = \Omega(n \log n)$.

Next, we consider the merging method originally proposed by Agarwal et al.~\cite{AEHW06} and mentioned in Section~\ref{sec:dyntrees}: to do a merge, insert the nodes of the shorter merge path one-by-one into the longer merge path.  We shall show that in the absence of cuts and if all merges are of leaves, the total number of nodes on the shorter of each pair of merge paths is $\Theta(n^{3/2})$, thus showing that this method of merging does not give a polylogarithmic amortized time bound for merging, though it does give a sublinear bound.  We denote by $p_i$ the number of nodes on the shorter of the merge paths in the $i^\mathrm{th}$ merge.

\begin{figure*}
\addtolength{\abovecaptionskip}{-.5cm}
\begin{center}
\resizebox{0.30\textwidth}{!} {\includegraphics{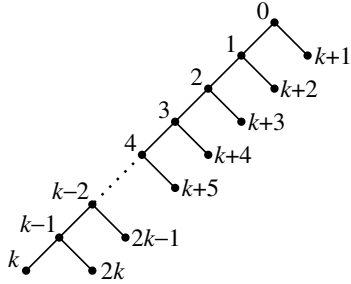}}
\end{center}
\caption{\label{fig:nsqrtn}
Initial tree for a sequence of merges whose shorter merge paths have $\Omega(n^{3/2})$ nodes.  The $i^\mathrm{th}$ merge is of the shallowest leaf and the leaf that is $\sqrt{k}$ deeper.}
\end{figure*}

To obtain the lower bound, start with the tree in Figure~\ref{fig:nsqrtn}, where $n = 2k + 1$ and $k$ is an arbitrary perfect square.  Do a sequence of $k - \sqrt{k}$ merges in which the $i^\mathrm{th}$ merge is $\merge(k + i, k + \sqrt{k} + i)$.  Each merge is of the shallowest leaf with the leaf that is $\sqrt{k}$ deeper.  For each merge, the longer merge path is the one starting from the deeper leaf; it contains $\sqrt{k} + 2$ nodes.  The shorter merge path contains two nodes for the first $\sqrt{k}$ merges, three for the next $\sqrt{k}$, four for the next $\sqrt{k}$, and so on.  Thus
\vspace{-.1cm}
\[
\sum_{i=1}^{k - \sqrt{k}} \hspace{-.1cm} p_i
= \sum_{i = 1}^{\sqrt{k} - 1} (i+1) \sqrt{k}
= \sqrt{k} \sum_{i = 1}^{\sqrt{k} - 1} \hspace{-.1cm} (i+1)
= \frac{k\sqrt{k} + k - 2\sqrt{k}}{2}
= \Omega(n^{3/2}).
\]

To show that this bound is tight to within a constant factor, assume without loss of generality that all insert operations precede all other operations.  Considering only nodes that participate in merge operations, let $\Phi$ be the number of unrelated pairs of nodes in the forest of mergeable trees.  After all insertions but before any merges, $\Phi = {n \choose 2}$.  As merges take place, $\Phi$ cannot increase but must remain non-negative.  The $i^\mathrm{th}$ merge decreases $\Phi$ by at least $(p_i - 1)^2$.  Thus $\sum_i (p_i - 1)^2 \le {n \choose 2}$.  Subject to this constraint, the sum of $p_i$'s is maximized when they are all equal, say to $p + 1$.  Then $mp^2 \le n^2$, which implies $p \le n/\sqrt{m}$ and $\sum_i p_i \le m + n \sqrt{m}$.  Since all merges are of leaves, $m < n$, giving $\sum_i p_i = O(n^{3/2})$.

Finally, we discuss lower bounds for three versions of the mergeable trees problem.  If cuts are allowed, the lower bound of P\u{a}tra\c{s}cu and Demaine~\cite{loglb:pd06} for the dynamic trees problem applies.  They show that in the cell probe model of computation, a sequence of intermixed insert, link, cut, and root operations take $\Omega(\log n)$ amortized time per operation.  This bound applies to mergeable trees even if there are no merge operations.  The data structure of Section \ref{sec:dyntrees} meets this bound except for merges, for which there is a logarithmic gap.  We conjecture that there is a solution to the mergeable trees problem with an amortized logarithmic bound for all operations; specifically, we think that the structure of Section \ref{sec:dyntrees} implemented using Sleator and Tarjan's self-adjusting dynamic trees~\cite{ST85} attains this bound.  We leave this question as the most interesting open problem emerging from our work.

If there are no cuts, we can obtain a lower bound by reducing sorting to the mergeable trees problem.  Specifically, we can sort $n$ numbers using a sequence of insert, merge, and parent operations, as follows.  We construct a one-node tree out of the first number by an insert.  For each successive number, we first construct a one-node tree by an insert and then merge it with the existing tree, which is a path.  We keep track of the maximum in this tree and use it as one parameter of the merge so that the new tree is also a path.  Finally we retrieve the numbers in reverse sorted order by starting at the maximum and doing $n-1$ parent queries.  Thus any data structure that supports insert, merge, and parent needs $\Omega(\log n)$ amortized time per operation, in any computation model in which sorting takes $\Omega(n \log n)$ time, such as a binary decision model.  In such models the structures of Sections \ref{sec:part-rank} and \ref{sec:weak} are optimum to within a constant factor.

In the absence of both cuts and parent queries, we can obtain a non-constant lower bound by reducing a form of disjoint set union to the mergeable trees problem.  The \emph{Boolean union-find} problem is that of maintaining a set of $n$ sets, initially singletons, under an intermixed sequence of two kinds of operations: $\mathit{unite}(A, B)$, which adds all elements in set $A$ to set $B$, destroying set $B$, and $\mathit{find}(x, A)$, which returns true if $x$ is in set $A$ and false otherwise.  Kaplan et al.~\cite{KST02} showed that a sequence of $m$ finds and intermixed unites takes $\Omega(m \alpha(m, n))$ time in the cell probe model with cells of size $\lg n$, where $\alpha$ is an inverse of Ackermann's function.  To solve the Boolean union-find problem using mergeable trees, we maintain for each set a tree, whose nodes are its elements and that is a path.  As the set identifier we use the maximum element in the set (with respect to an arbitrary total order); we can use an array to maintain the mapping from the names used by the set operations to the corresponding maximum nodes.  Initialization takes $n$ insert operations.  Each unite becomes a merge (of two different trees).  Each find can be done either by a single nca query or by two root queries.  The Kaplan et al.\ bound implies an $\Omega(\alpha(m, n))$ amortized time bound per mergeable tree operation for any structure that supports insert, merge, and either nca or root, for the cell probe model.  We conjecture that this lower bound is far from tight.

\vspace{.2cm}
\noindent{\emph{Acknowledgement.}} We thank Herbert Edelsbrunner for posing the mergeable trees problem and for sharing the preliminary journal version of~\cite{AEHW04} with us.

\bibliographystyle{plain}
\bibliography{rfw,ltg}

\end{document}